\documentstyle[12pt, twoside]{article}
   \newcommand {\nc}{\newcommand}
   \nc{\eq}{\begin{equation}}
   \nc{\en}{\end{equation}}
   \nc{\eqa}{\begin{eqnarray}}
   \nc{\ena}{\end{eqnarray}}
   \nc{\eqann}{\begin{eqnarray*}}
   \nc{\enann}{\end{eqnarray*}}
   \def\prf{{\bf{Proof:}}\\}
   \def\endprf{${\bf{\Box}}$\\}
   \newtheorem{definition}{Definition}
   \newtheorem{lemma}{Lemma}
   \newtheorem{theorem}{Theorem}
   \newtheorem{proposition}{Proposition}
   \newtheorem{corollary}{Corollary}

   \nc {\dfn}[1]{{\it{#1}}}
   \nc{\nn}{\nonumber}
   \def\dlt{\delta}
   \def\ep{\epsilon}
   \def\lam{\lambda}
   \def\thv{theory }
   \def\gp{group}
   \def\gpv{group }

   \def\subg{sub\gp}
   \def\subgv{sub\gpv }
   
   \def\sbv{sub\gpv }
   
   \def\nsgv{normal \subgv}
   
   \def\stbsbv{stationary \sbv}

   \def\symgv{symmetry \gpv}
   \def\ltgp{little \gp}
   \def\ltgpv{little \gpv}

   \def\rep{representation}
   \def\repv{representation }
   \def\Rep{Representation}
   \def\Repv{Representation }
   
   \def\repthv{\repv \thv}
   \def\indr{induced \rep}

   \def\prjrepv{projective \repv}
   \def\svrep{single-valued \rep}
   \def\svrepv{single-valued \repv}
   \def\tvrep{two-valued \rep}
   \def\tvrepv{two-valued \repv}
   \def\spinrep{spinor \rep}
   \def\spinrepv{spinor \repv}
   \def\md{module}
   \def\mdv{module }
   \def\irr{irreducible}
   \def\irrv{irreducible }
   \def\id{equivalent}
   \def\idv{equivalent }
   \def\alrep{inequivalent irreducible \rep s}
   \def\alrepv{inequivalent irreducible \rep s }
   \def\conj{conjugate}
   \def\conv{conjugate }
   
   \def\smdpv{semi-direct product }
   \def\dim{dimension}
   \def\diml{\dim al}
   \def\dimv{dimension }
   \def\dimlv{\dim al }
   
   \def\fdimlv{four-\dimlv}
   \def\fdim{four \dim}
   \def\fdimv{four \dimv}

   \def\strv{structure }
   \def\hep{high energy physics}

   \def\wpv{wreath product }
   \def\etal{{\it et al} }

   \def\decpv{decomposition }
   \def\clsf{classification}
   \def\clsfv{classification }

   \nc{\sqt}{\sqrt{2}}
   \nc{\tsqt}{2\sqt}
   \nc{\msqt}{$\sqt$}
   \nc{\nsqt}{$-\sqt$}
   \nc{\mtsqt}{$\tsqt$}
   \nc{\ntsqt}{$-\tsqt$}

   \def\ot{\otimes}
   
   \def\smdp{>\hspace{-0.2cm}\lhd}
   \def\otl{\bigotimes\limits}
   \def\prodl{\prod\limits}
   \def\suml{\sum\limits}
   \nc {\inv}[1]{#1^{-1}}
   \nc {\hc}[1]{{#1}^\dag}
   \nc {\cc}[1]{{#1}^\ast}
   \nc {\ad}[2]{Ad_{#1}({#2})}
   \nc {\wad}{\widetilde{Ad}}
   \nc {\wadf}[2]{\widetilde{Ad}_{#1}({#2})}
   \nc {\pb}[1]{{#1}^\ast}
   \def\tld{\tilde}
   \def\pr{\prime}
   \def\ov{\overline}
   \def\indxst{{\cal I}}
   
   \def\intg{{\cal Z}}
   \def\real{{\cal R}}
   \def\complex{{\cal C}}
   \def\quaternion{{\bf{H}}}

   \nc {\ga}[2]{{#1}[{#2}]}
   \nc {\cga}[1]{\ga{\complex}{#1}}
   \nc {\cgag}{\cga{G}}
   \nc {\eu}[1]{E^{#1}}
   \nc {\euf}{\eu{4}}
   \nc {\eun}{\eu{n}}
   \nc {\zn}[1]{\intg^{#1}}
   \nc {\zf}{\zn{4}}
   \nc {\zt} {Z_2}
   \nc {\ztn}[1]{\zt^{#1}}
   \nc {\ztt} {\ztn{2}}
   \nc {\ztth} {\ztn{3}}
   \nc {\ztf} {\ztn{4}}
   \nc {\ztnf}{\ztn{n}}
   \nc {\per}[1]{S_{#1}}
   \nc {\pern}{\per{n}}
   \nc {\irrr}[2]{IRR_{#1}({#2})}
   \nc {\irrc}[1]{\irrr{\complex}{#1}}
   \nc {\irrcg}{\irrc{G}}
   \def\hf{{1\over 2}}
   \def\ebr{\bar{e}}

   \nc {\mn}{(-)}
   \nc {\mns}[1]{\mn^{#1}}
   \nc {\mo}{(-1)}
   \nc {\mos}[1]{\mo^{#1}}
   \nc {\unit} {{\bf 1}}
   \nc {\unt} {\unit_{2\times 2}}
   \nc {\unth} {\unit_{3\times 3}}
   \nc {\unf} {\unit_{4\times 4}}
   \nc {\une} {\unit_{8\times 8}}
   \nc {\unw} {\unit_{12\times 12}}
   \nc {\zrt} {{\bf 0}_{2\times 2}}
   \nc {\zrth} {{\bf 0}_{3\times 3}}
   \nc {\zrf} {{\bf 0}_{4\times 4}}
    \def\CDalign#1{\bgroup\vcenter\bgroup\tabskip 2pt 
      \baselineskip 14pt \lineskip 3pt \lineskiplimit 3pt
      \halign\bgroup &\hfill$##$\hfill\crcr
      #1\crcr\egroup\egroup\egroup} 
    \def\CDdown{\Big\downarrow}       
    \def\CDrlabel#1{\vcenter{\hbox to0pt{$\scriptstyle#1$\hss}}} 
    \def\CDto{\mathop{\relbar\joinrel\longrightarrow}\limits}    
    \def\CDup{\Big\uparrow}  
    \def\CDeq{\Big\|}
  \nc{\act}[3]{S_{{#1}}^{{#2}}[{#3}]}
  \nc{\cact}[2]{S_{Cl}^{{#1}}[{#2}]}
  
   \nc {\prj}[4]{\pi_{#1,#2}#2\ot_{S_#1}e^o_{#3,#4}}
   \nc {\prjf}[2]{\prj{o}{#1}{\eta}{#2}}
   \nc {\prjff}{\prjf{h}{i}}
   \nc {\Prj}[4]{\Pi_{#1,#2;#3,#4}}
   \nc {\Prjf}[2]{\Prj{o}{#1}{\eta}{#2}}
   \nc {\Prjff}{\Prjf{h}{i}}
   \nc {\orbt}[2]{\pi_{#1,#2}}
   \nc {\orbtf}[1]{\orbt{o}{#1}}
   \nc {\orbte}{\orbtf{e}}
   \nc {\rp}[3]{\orbt{#1}{#2}(#3)}
   \nc {\rpf}[2]{\rp{o}{#1}{#2}}
   \nc {\stbrp}[2]{D^o_#1(\tld{s}(#2))}
   \nc {\stbrpme}[4]{\stbrp{#1}{#2}^{#3}_{#4}}
   \nc {\stbrpf}[1]{\stbrp{\eta}{#1}}
   \nc {\stbrpmef}[3]{\stbrpme{\eta}{#1}{#2}{#3}}
   \nc {\ch}[2]{\chi_{#1;#2}}
   \nc {\che}[3]{\ch{#1}{#2}(#3)}
   \nc {\chf}[1]{\ch{o}{#1}}
   \nc {\chff}{\chf{\eta}}
   \nc {\chef}[2]{\che{o}{#1}{#2}}
   \nc {\cheff}[1]{\chef{\eta}{#1}}
   \nc {\chsb}[1]{\chi^o_{#1}}
   \nc {\chsbe}[2]{\chsb{#1}(#2)}
   \nc {\chsbf}{\chsb{\eta}}
   \nc {\chsbef}[1]{\chsbe{\eta}{#1}}
   \nc {\cg}[1]{O_{#1}}
   \nc {\cgn} {\cg{n}}
   \nc {\oh} {\cg{4}}
   \nc {\ohd} {{\overline{\oh}}}
   \nc {\ztfb} {\overline{\ztf}}
   \nc {\iso}[2]{ISO_{#1}(#2)}
   \nc {\isod}[1]{\iso{d}{#1}}
   \nc {\eg}[1]{ISO(#1)}
   \nc {\egn}{\eg{n}}
   \nc {\fs}[1]{F(#1)}
   \nc {\fsf}{\fs{\emb}}
   \nc {\fx}[1]{I(#1)}
   \nc {\fxf}{\fx{\emb}}
   \nc {\wrn}[1]{\ztn{#1}\smdp\per{#1}}
   \nc {\wn}{\wrn{n}}
   \nc {\wnn}{\pern^{\zt}}
   \nc {\w}[1]{\per{#1}^{\zt}}
   \nc {\fix}[1]{\per{(n-#1)}\ot \per{p}}
   \nc {\fixp}{\fix{p}}
   \nc {\cb}[1]{C_{#1}}
   \nc {\cn}{\cb{n}}

   \def\cbgv{cubic \gpv}
   \nc {\prm}{\sigma}
   \nc {\prmt}{\tld{\prm}}
   \nc {\prmp}{\prm^\pr}
   \nc {\cyc}[2]{\tau_{{#1}{#2}}}
   \nc {\emb}{\iota}
   \nc {\epy}{\tld{\ep}}
   \nc {\de}[1]{d_{E^{#1}}}
   \nc {\den}{\de{n}}
   \nc {\dy}{\tld{d}}
   \nc {\repw}[2]{e_{(#1)#2}}
   \nc {\repww}[4]{\repw{#1}{#2}\ot\repw{#3}{#4}}
   \nc {\prw}[6]{\pi_{#1,#2}#2\ot_{F_#1}(\repww{#3}{#4}{#5}{#6})}
   \nc {\dm}[1]{d_{({#1})}}
   \nc {\sls}{{\it slash}}
   \nc {\slsv}{{\it slash} }
   \nc {\lcl} {{\it local}}
   \nc {\lclv} {{\it local} }
   \nc {\drln}[5]
    {\put(#1,#2){\line(#3,#4){#5}}}
   \nc {\ybxa}[4]
    {
     \begin{picture}(40,10)
      \drln {0}{0}{0}{1}{10}
      \drln {0}{0}{1}{0}{40}
      \drln {10}{0}{0}{1}{10}
      \drln {0}{10}{1}{0}{40}
      \drln {20}{0}{0}{1}{10}
      \drln {30}{0}{0}{1}{10}
      \drln {40}{0}{0}{1}{10}
      \drln {0}{0}{1}{1}{#1}
      \drln {10}{0}{1}{1}{#2}
      \drln {20}{0}{1}{1}{#3}
      \drln {30}{0}{1}{1}{#4}
     \end{picture}
    }
   \nc {\ybxb}[4]
    {
     \begin{picture}(30,20)
      \put(0,0){\line(0,1){20}}
      \put(0,0){\line(1,0){10}}
      \put(10,0){\line(0,1){20}}
      \put(0,10){\line(1,0){30}}
      \put(0,20){\line(1,0){30}}
      \put(20,10){\line(0,1){10}}
      \put(30,10){\line(0,1){10}}
      \drln {0}{0}{1}{1}{#1}
      \drln {0}{10}{1}{1}{#2}
      \drln {10}{10}{1}{1}{#3}
      \drln {20}{10}{1}{1}{#4}
     \end{picture}
    }
   \nc {\ybxc}[4]
    {
     \begin{picture}(20,20)
      \put(0,0){\line(0,1){20}}
      \put(0,0){\line(1,0){20}}
      \put(10,0){\line(0,1){20}}
      \put(0,10){\line(1,0){20}}
      \put(0,20){\line(1,0){20}}
      \put(20,0){\line(0,1){20}}
      \drln {0}{0}{1}{1}{#1}
      \drln {10}{0}{1}{1}{#2}
      \drln {0}{10}{1}{1}{#3}
      \drln {10}{10}{1}{1}{#4}
     \end{picture}
    }
   \nc {\ybxd}[4]
    {
     \begin{picture}(20,30)
      \drln {0}{0}{0}{1}{30}
      \drln {0}{0}{1}{0}{10}
      \drln {10}{0}{0}{1}{30}
      \drln {0}{10}{1}{0}{10}
      \drln {0}{20}{1}{0}{20}
      \drln {0}{30}{1}{0}{20}
      \drln {20}{20}{0}{1}{10}
      \drln {0}{0}{1}{1}{#1}
      \drln {0}{10}{1}{1}{#2}
      \drln {0}{20}{1}{1}{#3}
      \drln {10}{20}{1}{1}{#4}
     \end{picture}
    }
   \nc {\ybxe}[4]
    {
     \begin{picture}(10,40)
      \drln {0}{0}{0}{1}{40}
      \drln {10}{0}{0}{1}{40}
      \drln {0}{0}{1}{0}{10}
      \drln {0}{10}{1}{0}{10}
      \drln {0}{20}{1}{0}{10}
      \drln {0}{30}{1}{0}{10}
      \drln {0}{40}{1}{0}{10}
      \drln {0}{0}{1}{1}{#1}
      \drln {0}{10}{1}{1}{#2}
      \drln {0}{20}{1}{1}{#3}
      \drln {0}{30}{1}{1}{#4}
     \end{picture}
    }
   \def\orb{{\bf{Orbit }}}
   \nc{\op}{orientation-preserved}
   \nc{\opv}{orientation-preserved }
   \nc{\on}[1]{SO_{#1}}
   \nc{\ohn}[1]{O_{#1}}
   \nc{\of}{\on{4}}
   \nc{\ohf}{\ohn{4}}
   \nc{\ofd}{\overline{\of}}
   \nc{\ohfd}{\overline{\ohf}}
   \nc{\cir}[1]{e^{i({#1})\pi}}

   \nc {\norm}[1]{\parallel{#1}\parallel}

   \nc{\groupsum}[1]{{\sum_{#1}}^{\pr}}
   \nc{\reducedsum}[1]{{\sum_{#1}}^{\pr\pr}}
 \title{Structure and representation theory for the double group of the four-dimensional cubic group}
 \author{Jian Dai$^\dag$,
  Xing-Chang Song\\
  Theory Group, Department of Physics\\
  Peking University, Beijing, 100871, P. R. China\\
  \dag Room 2082, Building48,
  Peking University, Beijing, 100871, P. R. China\\
  {\bf E-mails:} daijianium@yeah.net,
  songxc@ibm320h.phy.pku.edu.cn
  }
 \date{February 1st, 2001}
 
\begin{document}
  \pagestyle{myheadings}
  \markright{Structure and Representations for $\overline{O_4}$}
  \begin{titlepage}
  \maketitle
  \begin{abstract}
  \noindent
   Hyper-cubic groups in any \dimv are defined and their \conv classifications and representation
   theories are derived. Double \gpv and \spinrepv are introduced.
   A detailed calculation is carried out on the structures of four-\dimlv cubic group
   $\oh$ and its double group, as well as all in\idv \svrep s and \spinrep s of $\oh$.
   All \rep s are derived adopting Clifford theory of
   \decpv of \indr s. Based on these results, single-valued and spinor \rep s of
   the \opv \sbv of $\ohf$ are calculated.\\


  \end{abstract}
  \end{titlepage}
  \section{Introduction}
   It is well-known that electrons stay
   in \spinrep s of the \symgv of a given lattice in condensed matter physics;
   it is reasonable to assume that quarks, leptons, as well as baryons, should
   reside in {\it \spinrep s} of the \symgv of a \fdimlv lattice in
   lattice field theory (the concept of
   ``\spinrep'' will be clarified in the next section). Accordingly, to explore the
   \strv and \rep s (\spinrep s especially) of such groups has important significance in \hep.\\
   \\
   In this paper, we concentrate on the case of hyper-cubic lattices, though
   they are not the maximum symmetric lattices in \fdimv \cite{bc}. In history, the first \rep-theoretical
   consideration of \symgv of such lattices was given by A.
   Young \cite{young}. Then mathematicians worked in this field
   due to the interest of \wpv \cite{sp1}\cite{sp2} to which A.
   Kerber gave a thorough review in his book \cite{kerber}.
   Physicists took part in after K. G. Wilson introduced
   lattice gauge theory \cite{wilson}. M. Baake \etal first gave an explicit description of characters
   of \fdimlv \cbgv \cite{baake}; J. E. Mandula \etal derived the same results using a different method
   \cite{mandula}. As for \spinrep s, Mandula \etal resolved this problem for what we call {\it
   orientation-preserved} four-dimensional cubic-group in \cite{m2}.\\
   \\
   In this paper, the power of Clifford theory on \decpv of \indr s (Sec.\ref{indrep}) is fully applied.
   A systematic and schematic description of \conv \clsfv and \repthv of generalized
   \cbgv $\cgn$, as well as the concept of orientation-preserved subgroup of them $SO_n$,
   is given in Sec.\ref{ohstr}. Double group is introduced in Sec.\ref{doublegroup}
   to clarify the terms ``\svrep" and ``\tvrep (\spinrep)". Then
   specifying these general results to \fdim, we give a detailed description of \strv and \conv \clsfv of $\oh$ (Sec.\ref{sec.str}),
   its double $\ohd$ (Sec.\ref{ohdstr}), and those of $SO_4$, $\overline{SO_4}$ (Sec.\ref{so4case}).
   We derive all in\idv
   \svrep s as well as \spinrep s of $\oh$, adopting Clifford theory
   (Sec.\ref{repofohd}). Based on these results, we reproduce
   \repthv of $\overline{SO_4}$ in Sec.\ref{so4rep}.\\

   It should be pointed out that the ``spinor" part for $O_4$ of our work is
   completely new and that although other results are well-known,
   our method to derive them is much more tidy and systematic than
   that used by other authors who gave the same results, thanks for the power of
   Clifford theory.
  \section{Conceptual foundations}
   \subsection{Clifford theory on decomposition of \indr s}
   \label{indrep}
    Two results of Clifford theory, a powerful method for decomposing \indr s of a given \gpv $E$ with a \nsgv $N$
    \cite{cr}\cite{s}\cite{cs}\cite{qiu}, will be applied in this paper. We will use $\cga{E}$ for the group algebra of $E$ in complex
    field and $G$ for $E/N$ below. The first result is
    \begin{theorem}(Clifford)\cite{clif}\cite{cr}
    \label{clifford}
     Let $M$ be a simple $\cga{E}$-\md, and $L$ a simple
     $\cga{N}$-sub\mdv of $M_N$ s.t. $L$ is stable relative to
     $E$, i.e. $L$ is isomorphic to all of its \conj s. Then
     \[
      M\cong L\ot_{\complex}I
     \]
     for a left ideal $I$ in $End_{\cga{E}}L^E$. The $E$-action
     on $L\ot_{\complex}I$ is given by
     \[
      x\mapsto U(x)\ot V(x), x\in E
     \]
     where $U:E\rightarrow GL(L)$ is a \prjrepv of $E$ on $L$, and
     $V:E\rightarrow GL(I)$ is a \prjrepv of $G$, that is, $V(x)$
     depends only on the coset $xN$ of $x$ in $G$, for each $x\in
     E$. The factor sets associated with $U$ and $V$ are inverse
     of each other.
    \end{theorem}
    The second result can be regarded as a special case of Theorem
    \ref{clifford}. Let $E=N\smdp G, |E|<\infty$ and $N$ be abelian, then adjoint action
    of $G$ upon $N$ makes $N$ a $G$-module. This $G$-action can be extended naturally
    to a $G$-action upon $\cga{N}$ by linearity.
    Define $\Pi(N):=\{\pi_\mu\}\subset \cga{N}$,
    \eq
    \label{defpro}
     \pi_\mu:=\sum_{a\in N}{\chi_\mu (\inv{a}) a}
    \en
    where $\chi_\mu$ are all \irrv \rep s of
    $N$. The $G$-action on $\Pi(N)$ is closed and thus $\Pi(N)$
    is separated into orbits
    $\Pi(N)=\coprod\limits_{o\in \indxst}{\Pi_o}$ where $\indxst$ is a index set to label
    different orbits. For each $\Pi_o$,
    choose one of its element and denote it as $\orbte$. The
    stablizer of each $\orbte$ in $G$ (\dfn{\ltgp}) is denoted as $S_o$.
    There is a bijection from $G/S_o=\{hS_o \}$ to $\Pi_o$ defined
    by
    \eq
    \label{orbit}
     Ad_h(\orbte)=h\orbte \inv{h}=:\orbtf{h}
    \en
    where $\{h\}$ is a system of representatives of left cosets
    $G/S_o$. Define
    \eq
    \label{main}
     \Prjff\equiv\prjff
    \en
    in which $\{e^o_{\eta,i}|i=1,2,...,d^o_\eta\}$ with fixed $o,\eta$ and $\eta$ is the $\eta$th \irrv \repv of
    $S_o$ whose \dimv is $d^o_\eta$, then
    \begin{proposition}(\ltgpv method)\cite{s}\cite{cs}\cite{qiu}
    \label{rfm}
     \begin{enumerate}
      \item For each fixed $(o,\eta)$, $\{\Prjff\}$ induces an \irrv \repv of
       $E$, denoted as $D_{o,\eta}$;
      \item If $(o,\eta)\neq (o^\pr,\eta^\pr)$, then
       $D_{o,\eta}$ and $D_{o^\pr,\eta^\pr}$ are
       in\id;
      \item $\{D_{o,\eta}\}$ gives all in\idv \irrv \rep s of $E$.
     \end{enumerate}
    \end{proposition}
   \subsection{Cubic group in any dimension}
   \label{ohstr}
    The symmetry group of a cube including inversions in three dimensional Euclidean space,
    which is denoted as $O_h$ in the theory of point groups \cite{hs},
    can be generalized into any $n$-dimensional Euclidean space
    $\eun$, along two different approaches whose results are \id.
    The first approach of generalization which is very natural and straightforward
    is geometrical. An \dfn{$n$-cube} (or \dfn{hyper-cube in $\eun$}) $\cn$ is defined to
    be a subset of $\eun$, $\cn=\{p|x^i(p)=\pm 1\}$, where $x^i:\eun\rightarrow \real, i=1,2,...,n$
    are coordinate functions of $\eun$, together with the distance inherited from $\eun$.
    \dfn{$n$-Cubic group} (\dfn{hyper-cubic group of degree $n$}) $O_n$ consists of all isometries of $\eun$ which stabilize
    $\cn$. While the second approach of generalization is
    algebraic. $O_h$ has a \smdpv structure
    as $\ztth\smdp\per{3}$ \cite{s}; we generalize this to
    $\wn$ which is just a \dfn{wreath product} $\zt\wr\pern$ of
    $\zt$ with $\pern$. We point out that these two
    generalizations are identical. Let $\{e_i\}$ be
    a standard orthogonal basis of $\eun$, namely $x^j(e_i)=\dlt^j_i$.
    Define $n+1$ points in $\cn$ to be $p_0=(-1,-1,...,-1),
    p_i=p_0+2e_i$.
    \begin{lemma}
     $\forall \ep \in \cgn$, $\ep$ is entirely determined by
     images $\ep(p_i), i=0,1,2,...,n$.
    \end{lemma}
    \prf
     The fact that $\ep$ is an isometry of $\eun$ ensures the equality of Euclidean distances $d(p,p_i)=d(\ep(p),\ep(p_i))$,
     $i=0,1,...,n$
     for any other $p$ in $\cn$.
     If all $\ep(p_i)$ are given, $\ep(p)$ will be fixed for any other $p$
     accordingly due to the fundamental lemma of Euclidean geometry (lemma
     \ref{fundament} in Appendix \ref{app2}). In fact, the existence of solution in lemma \ref{fundament}
     is guaranteed by that $\ep$ stablizes $\cn$ and
     lemma \ref{fundament} itself ensures the uniqueness.\\
    \endprf
    To fix $\ep(p_0)$, there are $2^n$ ways; while for a fixed $\ep(p_0)$, there are $n!$
    possibilities to fix $\ep(p_i), i=1,2,...,n$. Therefore, $|\cgn|=2^n\cdot n!$.
    \begin{proposition}(Structure of $\cgn$)
    \label{On}
     \eq
     \label{Oneq}
      \cgn\cong \zt^n\smdp S_n
     \en
    \end{proposition}
    \prf
     Introduce a class of isometries in $\eun$:
     \eq
     \label{cong}
      \prm(e_i)=e_{\prm(i)}; I_i(e_j)=(1-2\dlt_{ij})e_j,
      i=1,2,...,n
     \en
     where $\prm\in \pern$ permutes the axes and $I_i$ inverts the $i$th
     axis. Subjected to the relations
     \eq
     \label{genr}
      I_i^2=e, I_iI_j=I_jI_i, i,j=1,2,...,n;
      \prm I_i=I_{\prm(i)}\prm, \prm\in\pern
     \en
     these isometries generate a sub-group of $\cn$ isomorphic to $\wn$ whose order
     is $2^n n!=|\cgn|$. So (\ref{Oneq}) follows.\\
    \endprf
    A. Kerber gave a detailed introduction on the \conv classification and \repthv
    of a general \wpv $N\wr G$ in \cite{kerber}. We specify his general results to our case $\wn\cong \zt\wr \pern$.\\
    \\
    Some fundamental facts about symmetrical \gp s $\pern$ should be recalled \cite{hs}.
    Each element $\prm\in \pern$ has a cycle decomposition
    \eq
    \label{prmdc}
     \prm=
     \left(
       \begin{array}{cccc}
        1      &2      &\ldots&n      \\
        \prm(1)&\prm(2)&\ldots&\prm(n)
       \end{array}
      \right)
      =\prodl_{k=1}^n\prodl_{\alpha=1}^{\nu_k}\cyc{k}{\alpha}
    \en
    where $\cyc{k}{\alpha}$ are independent $k$-cycles, which can be expressed as $(a_1a_2...a_k)$, and write $n(k,\alpha)
    =\{a_1,a_2,...,a_n\}$. The cycle structure of $\prm$ can be
    represented formally as
    \eq
    \label{sh}
     (\nu)=\prodl_{k=1}^n{(k^{\nu_{k}})}
    \en
    where $\{\nu_k\}$ satisfies $\suml_{k=1}^n{k\cdot \nu_k}=n$.
    Two elements in $\pern$ are \conv \id, iff they have the same cycle
    structure. The number of elements in class
    $(\nu)$ is equal to $N_{(\nu)}=n!/\prodl_{k=1}^n{(k^{\nu_k}\nu_k!)}$.
    Each cycle structure $(\nu)$ can be visualized by one unique Young diagram which is
    denoted also by $(\nu)$.
    There is a one-one correspondence between all in\idv \irrv
    \rep s of $\pern$ and all Young diagrams, which enable us to represent each \irrv \repv
    by the corresponding Young diagram $(\nu)$. We write the basis of one of these \rep s $(\nu)$ in $\dm{\nu}$ \dimv
    as $\repw{\nu}{i}, i=1,2,...,\dm{\nu}$.
   \\
   \\
    We point out that the \conv \clsfv of $\cgn$ has a deep relation with that of
    $\pern$. A generic element in $\zt\wr\pern$ can be written as
    \eq
    \label{cycle}
     \prm \cdot\prodl_i{I_i^{s_i}}=
     \left(
       \begin{array}{cccc}
        1                 &2                 &\ldots&n                \\
        \mns{s_1}\prm(1)&\mns{s_2}\prm(2)&\ldots&\mns{s_n}\prm(n)
       \end{array}
     \right)
    \en
    in which $s_i\in\intg/2\intg$. We call the r.h.s. of (\ref{cycle}) by
    {\it permutation with signature}.
    $\prm\prodl_i{I_i^{s_i}}$ can be
    decomposed according to (\ref{prmdc}), i.e.$\prodl_i{I_i^{s_i}}=\prodl_{k=1}^n\prodl_{\alpha=1}^{\nu_i}\prodl_{a\in n(k,\alpha)}
    I_a^{s_a}$ and
    \eq
    \label{decompwnn}
     \prm\prodl_i{I_i^{s_i}}=\prodl_{k=1}^n\prodl_{\alpha=1}^{\nu_i}(\cyc{k}{\alpha}\prodl_{a\in n(k,\alpha)}
     I_a^{s_a})
    \en
    The {\it cycle with signature} is defined to be
    \[
    \label{cycle1}
     \cyc{k}{\alpha}\prodl_{a\in n(k,\alpha)}I_a^{s_a}=
     \left(
       \begin{array}{cccc}
        a_1              &a_2              &\ldots &a_k            \\
        \mns{s_{a_1}}a_2 &\mns{s_{a_2}}a_3 &\ldots &\mns{s_{a_k}}a_1
       \end{array}
      \right)
    \]
    For two independent $(k,\alpha),(k^\pr,\alpha^\pr)$, it is easy
    to verify that
    \[
     \cyc{k}{\alpha}\cyc{k^\pr}{\alpha^\pr}=\cyc{k^\pr}{\alpha^\pr}\cyc{k}{\alpha},
     \prodl_{a\in n(k,\alpha)}I_a^{s_a}\cyc{k^\pr}{\alpha^\pr}=\cyc{k^\pr}{\alpha^\pr}\prodl_{a\in
     n(k,\alpha)}I_a^{s_a},
     \prodl_{a\in n(k^\pr,\alpha^\pr)}I_a^{s_a}\cyc{k}{\alpha}=\cyc{k}{\alpha}\prodl_{a\in n(k^\pr,\alpha^\pr)}I_a^{s_a}
    \]
    \begin{proposition}
    \label{ccpro}
    \cite{sp1}\cite{sp2}\cite{kerber}We use $\sim$ to denote \conv
    \id.
     \begin{enumerate}
      \item (descent rule)
       \eq
       \label{des}
        \prm\prodl_i{I_i^{s_i}}\sim \prm^\pr\prodl_i{I_i^{s_i^\pr}}
        \Rightarrow\prm\sim\prm^\pr
       \en
      \item (permutation rule) Let
       \[
        \prmt=
         \left(
          \begin{array}{cccc}
           1       &2       &\ldots&n       \\
           \prmt(1)&\prmt(2)&\ldots&\prmt(n)
          \end{array}
         \right)
         =
         \left(
          \begin{array}{cccc}
           \prm (1)&\prm (2)&\ldots&\prm (n)\\
           \prmp(1)&\prmp(2)&\ldots&\prmp(n)
          \end{array}
         \right)
       \]
       then
       \eq
       \label{cyc}
        \prmt(\prm\prodl_i{I_i^{s_i}})\inv{\prmt}=
         \left(
          \begin{array}{cccc}
           \prmt(1)         &\prmt(2)         &\ldots&\prmt(n)        \\
           \mns{s_1}\prmp(1)&\mns{s_2}\prmp(2)&\ldots&\mns{s_n}\prmp(n)
          \end{array}
         \right)
       \en
      \item (signature rule within one cycle) Let $\cyc{k}{\alpha}$ be a $k$-cycle
       and $a_0$ be a given number in $n(k,\alpha)$, then
       \eq
       \label{signat}
        \cyc{k}{\alpha}\prodl_{a\in n(k,\alpha)}I_a^{s_a}\sim
        \cyc{k}{\alpha}\prodl_{a\in
        n(k,\alpha)}I_a^{s_a+\dlt_{aa_0}+\dlt_{a,\cyc{k}{\alpha}(a_0)}}
       \en
       Note that $\cyc{k}{\alpha}(a_0)$ is calculated modulo $k$ (the subscripts of $I_a$ are always understood in this way).
      \item (signature rule between two cycles) Let $\cyc{k}{\alpha},\cyc{k}{\beta}$
       be two independent $k$-cycles and we define a bijection $\theta:
       n(k,\alpha)\rightarrow n(k,\beta), a_i\mapsto b_i$. Then
       \eq
       \label{exchg}
        \cyc{k}{\alpha}\prodl_{a\in n(k,\alpha)}I_a^{s_a}\cdot\cyc{k}{\beta}\prodl_{b\in n(k,\beta)}I_b^{s_b}
        \sim
        \cyc{k}{\alpha}\prodl_{a\in n(k,\alpha)}I_a^{s_{\theta(a)}}\cdot\cyc{k}{\beta}\prodl_{b\in
        n(k,\beta)}I_b^{s_{\inv{\theta}(b)}}
       \en
     \end{enumerate}
    \end{proposition}
   This theorem ensures \conv \clsfv of $\zt\wr\pern$ is totally determined
   by the structure of cycles with signature. We verify this
   statement by generalizing Young diagram technology. First,
   draw a {\it Young diagram with numbers and signatures} for each element $\prm\prodl_i{I_i^{s_i}}\in\zt\wr\pern$
   according to the decomposition Eq.(\ref{decompwnn}) by the
   following rules:
   \begin{enumerate}
    \item
    \label{r1}
     Plot Young diagram of the class in $\pern$ to which $\prm$ belongs and
     fill each column of this Young diagram with numbers in corresponding cycle by cyclic ordering from up-most box to down-most box.
    \item Draw a \slsv in the Young box if the number in this box is mapped to a minus-signed number.
   \end{enumerate}
   Secondly, partition elements in $\zt\wr\pern$ by their cycle structure in $\pern$ and Eq.(\ref{des}) guarantees elements belong to
   different partitions can not be \conv \id. Eq.(\ref{cyc}) implies that all the numbers that we filled by rule \ref{r1} are
   unnecessary, so smear them out and leave boxes and \sls es only. Within each column, Eq.(\ref{signat}) says that the positions of \sls es
   make no difference. What's more, in fact only that the total
   number of \sls es is even or odd distinguishes different
   classes. Therefore we regulate each column to contain zero or one
   \slsv at the bottom box. Eq.(\ref{exchg}) shows that we can
   not distinguish the case that one column without any \slsv (Mr. Zero) is put to the left
   to another column with one \slsv (Mr. One) from that Mr.Zero is to the right of Mr.One, if they have same
   cyclic length; thus we regulate that Mr.Zero shall always
   stand left to Mr.One. Therefore, conjugate classes of $\zt\wr\pern$ can be uniquely characterized by generalizing Young
   diagrams containing \sls es. Following Eq.(\ref{sh}), we represent \conv classes by
   \eq
   \label{syt}
    (\nu^+,\nu^-)=\prodl_{k=1}^n{(k^{\nu_{k}^{+}+\nu_{k}^{-}})}
   \en
   where $\nu_k^+$ is the number of Mr.Zero-type $k$-cycles and
   $\nu_k^-$ is that of Mr.One-type $k$-cycles, which satisfy $\nu_{k}^{+}+\nu_{k}^{-}=\nu_k$. It is not difficult to
   check some numerical properties of \conv classes of $\zt\wr\pern$.
   \begin{corollary}
    \begin{enumerate}
     \item Given a class $(\nu)$ in $\pern$, there are
      \eq
      \label{n1}
       \prodl_{k=1}^n {(1+\nu_k)}
      \en
      classes in $\zt\wr\pern$ which descend
      to $(\nu)$.
     \item The number of elements in a class $(\nu^+,\nu^-)$ is
      \eq
      \label{n2}
       N_{(\nu^+,\nu^-)}=N_{(\nu)}
       \prodl_{k=1}^n (C_{\nu_k}^{\nu_k^+} (\suml_{i=0}^{[{k\over 2}]}C_k^{2i})^{\nu_k^+} (\suml_{j=1}^{[{{k+1}\over 2}]} C_{k}^{2j-1})^{\nu_k^-})
      \en
      where $C_m^n$ is combinatorial number defined to be $m!/(n!(m-n)!)$.
     \item The order of a class $(\nu^+,\nu^-)$ is
      \eq
      \label{n3}
       lcm(\{k\cdot 2^{\dlt(\nu_k^-)}|\nu_k\neq 0\})
      \en
      where $\dlt(\nu_k^-)=0$, if $\nu_k^-=0$; $\dlt(\nu_k^-)=1$, if
      $\nu_k^->0$.
     \item Determinant (signature, parity) of a class
      \eq
      \label{n4}
       det((\nu^+,\nu^-))=\mos{\suml_{k=1}^n{\nu_k^-}}\cdot
       det((\nu))
      \en
      where $det((\nu))$ is the determinant of $(\nu)$ in $\pern$.
    \end{enumerate}
   \end{corollary}
   All in\idv \irr \rep s of $\ztnf$ can be expressed as
   \eq
   \label{zrp}
    \chi_{(s)}:=\otl_{p=1}^n{\chi_{\mns{s_p}}}
   \en
   in which $s_p\in \intg/2\intg, p=1,2,...,n$ and
   $\chi_{\mn},\chi_{(+)}$ are two \irrv \rep s of $\zt$ with
   $\chi_{(+)}$ being the unit \rep. Thus $\pi_{(s)}$ can be
   defined by Eq.(\ref{defpro}) and $\Pi(\ztnf)=\{\pi_{(s)}\}$.
   Note that $\pi_{(s)}$ satisfy $\pi_{(s)}\pi_{(s^\pr)}=\pi_{(s\cdot
   s^\pr)}$ where $(s\cdot s^\pr)(p)=s(p)s^\pr(p)$.
   $\Pi(\ztnf)$ is divided into $n+1$ orbits under the
   $\pern$-action, namely
   $\Pi(\ztnf)=\coprod\limits_{p=0}^n{\Pi_p}$.
   For a given $p$, $\Pi_p$ consists of those $\pi_{(s)}$ who has $p$
   components in $(s)$ equal to $1$, other $n-p$ components equal
   to $0$; hence $|\Pi_p|=C_n^p$.
   Each $\orbt{p}{e}$ is specified to a $\pi_{(s)}$ with $s_p=0, p=1,2,...,n-p;
   s_p=1, p=n-p+1,...,n$, whose \stbsbv is just $\fixp$, denoted as
   $F_p$. Representatives of left-cosets in $S_n/F_p$ are written as $\prm_r$.  , then according
   to Eqs.(\ref{orbit})(\ref{main}) and Theorem \ref{rfm},
   \begin{proposition}(\Repv theory of $\zt\wr\pern$)
   \label{ron}
    \[
    \Prj{p}{\prm_r}{(\mu)i}{(\nu)j}=\prw{p}{\prm_r}{\mu}{i}{\nu}{j}
    \]
    give all in\idv \irrv \rep s of $\zt\wr\pern$ when $(p,(\mu),(\nu))$
    runs over its domain.
   \end{proposition}
   where $\orbt{p}{\prm_r}=\ad{\prm_r}{\orbt{p}{e}}$ whose $(s)$ will be denoted as
   $(s^{(p\prm_r)})$.
   \begin{corollary}
    \begin{enumerate}
     \item (Burside formula) $\suml_{(p,(\mu),(\nu))} (C_n^p \dm{\mu}\dm{\nu})^2=2^n n!$
     \item The number of \conv classes is
      $\suml_{(p,(\mu),(\nu))}1$.
     \item (\Repv matrix element)
      Given $\prm\prodl_q{I_q^{t_q}}\in \zt\wr\pern$,
      \[
       D_{(p,(\mu),(\nu))}(\prm\prodl_q{I_q^{t_q}})^{\prm_r^\pr i^\pr j^\pr}_{\prm_r
       ij}=\dlt^{\prm^\pr}_{\tld{\prm_r}(\prm\prm_r)} D_{(\mu)}(\prm_{(n-p)}(\prm\prm_r))^{i^\pr}_i
       D_{(\nu)}(\prm_p(\prm\prm_r))^{j^\pr}_j \prodl_q{\mns{s^{(p\prm_r)}_q t_q}}
      \]
     \item (Character)
      \[
       \chi_{(p,(\mu),(\nu))}(\prm\prodl_q{I_q^{t_q}})=\dlt^{\prm^\pr}_{\tld{\prm_r}(\prm\prm_r)}
       \chi_{(\mu)}(\prm_{(n-p)}(\prm\prm_r))
       \chi_{(\nu)}(\prm_p(\prm\prm_r))
       \prodl_q{\mns{s^{(p\prm_r)}_q t_q}}
      \]
    \end{enumerate}
   \end{corollary}
   where $\tld{\prm_r},\prm_{(n-p)},\prm_p$ map an element in
   $\pern$ to its decompositions according to $\pern/F_p$,
   $\per{n-p}$ and $\per{p}$ respectively.\\

   At the end of this subsection, we introduce the {\it orientation-preserved} n-cubic group $SO_n$ which
   is a normal subgroup of $O_n$
   \eq\label{son}
    SO_n:=(\cgn\cap SO(n))\lhd\cgn
   \en
   Define $\ztn{n}|_e$ as a \sbv of $\ztn{n}$ generated by $I_iI_j, i\neq
   j$ and $\ztn{n}|_o:=\ztn{n}\backslash \ztn{n}|_e$.
   Then
    \eq
    \label{decomOn}
     \on{n}=(\ztn{n}|_e\cdot A_{n})\bigsqcup (\ztn{n}|_o\cdot (S_{n}\backslash A_{n}))
    \en
    in which $\cdot$ is product of two subsets in a group, $A_{n}$
    stands for the alternative \sbv in $S_{n}$.
   Thus, $|\on{n}|=(2^{n}\cdot (n)!)/2$.
   \subsection{Double group and spinor \rep}
   \label{doublegroup}
    Some fundamental facts of Clifford algebra are necessary for giving the definition and properties of
    double groups.
    Denote the Clifford algebra upon Euclidean space $V$ as
    $Cl(V)$; the isometry $x\mapsto -x$ on $V$ extends to an automorphism
    of $Cl(V)$ denoted by $x\mapsto
    \tilde{x}$ and referred to as the canonical
    automorphism of $Cl(V)$.
    We use $Cl^\ast(V)$ to denote the multiplicative group of
    invertible elements in $Cl(V)$ and the Pin group is the subgroup of $Cl^\ast(V)$ generated by
    unit vectors in $V$ , i.e.
    \[
     Pin(V):=\{a\in Cl^\ast(V): a=u_1\cdots u_r, u_j\in V,
     \|u_j\|=1\}
    \]
    Proofs of the following four statements can be found in
    \cite{harvey}.
    \begin{lemma}
    \label{reflection}
     If $u\in V$ is nonnull, then $R_u$, reflection along $u$, is
     given in terms of Clifford multiplication by
     \[
      R_u x=-uxu^{-1}, \forall x\in V
     \]
    \end{lemma}
    \begin{theorem}
     The sequence
     \[
      0\rightarrow \zt\rightarrow
      Pin(V)\stackrel{\wad}{\rightarrow} O(V)\rightarrow 1
     \]
     is exact, in which
     \[
      \wadf{a}{x}:=\tilde{a}xa^{-1}, \forall x\in Cl(V),
      a\in Pin(V)
     \]
    \end{theorem}
    We will usually write $\wad$ just by $\pi$ as a surjective homomorphism.
    \begin{proposition}
    \label{rep}
     $Cl(E^4)$, as an associative algebra with unit, is isomorphic
     to $M_2(\quaternion)$ where $\quaternion$ denotes quaternions.
    \end{proposition}
    \begin{lemma}
    \label{su4}
     Under the above algebra isomorphism, the image of $Pin(E^4)$
     is a subset of $SU(4)$.
    \end{lemma}
    Now we give the main definition of this paper.
    \begin{definition}
     Let $\ep$ be an injective homomorphism from a \gpv $G$ to $O(n)$, then the double group or the spin-extension of $G$
     with respect to $\ep$ is defined to be $D_n(G,\ep):=\pi^{-1}(\ep(G))$.
    \end{definition}
    An introduction to double groups in three dimension can be
    found in \cite{joshi}.
    Following elementary facts in the theory of group extension \cite{brown}, this diagram
    \[
     \CDalign{0&\CDto&\zt  &\CDto            &Pin(E^n)                        &\CDto^{\pi}&O(n)&\CDto&1 \cr
               &     &\CDeq&                 &\CDup&                &\CDup\CDrlabel{\epsilon}&        \cr
              0&\CDto&\zt  &\CDto            &\pi^{-1}(\ep(G))                     &\CDto           &G                       &\CDto&1  }
    \]
    is commutative. If $\ep_1(G)\sim \ep_2(G)$, there is
    \[
     \CDalign{0&\CDto&\zt  &\CDto^i         &\pi^{-1}(\epsilon_1(G))&\CDto^{\pi}       &\epsilon_1(G)&\CDto&1 \cr
              &     &\CDeq&                &\CDdown                   &                  &\CDdown       &        \cr
             0&\CDto&\zt  &\CDto^{i^\prime}&\pi^{-1}(\epsilon_2(G))&\CDto^{\pi^\prime}&\epsilon_2(G)&\CDto&1  }
    \]
    Note that the double group is not a universal object for a
    given abstract group $G$ but a special type of $\zt$-central extension of $G$
    subjected to the embedding
    $\ep$. For example, the results of doubling two $\zt$ subgroups in
    $O(2)$, $I:=\{1, \sigma\}, R:=\{1, R(\pi)\}$ where $\sigma$ denotes
    reflection along $y$-axis and $R(\pi)$ is the rotation over $\pi$, are $\pi^{-1}(I)\cong \zt\ot \zt$ while
    $\pi^{-1}(R)\cong Z_4$. Nevertheless, we will use symbol $\bar{G}$ to denote the double group
    at most cases where $n$ and $\epsilon$ are fixed, and will not distinguish $G$ from $\ep(G)$. Meanwhile,
    symbol $\ebr$ is adopted to refer $-1$ in Clifford algebra and is called {\it{central element}}.\\
    \\
    Let $s: G\rightarrow \bar{G}, s.t. \pi s=Id_G$, namely $s$ is a cross-section of $\pi$.
    There is a property of the \conv classes of $\bar{G}$ which is
    easy to verify.
    \begin{lemma}
    \label{conjugat}
     Let $C$ be a \conv class in $G$, then either will $\pi^{-1}(C)$
     be one \conv class in $\bar{G}$ satisfying
     $\forall g\in C, s(g)\sim -s(g)$; or it will split into two \conv classes $C_1$, $C_2$ in
     $\bar{G}$ s.t. $\forall g\in C, s(g)\in C_1\Leftrightarrow -s(g)\in C_2$.
    \end{lemma}
    We will give a more deep result on the splitting of \conv
    classes when doubling $G$ to $\bar{G}$ in our another paper.\\

    Let $r$ be an \irrv \repv of $\bar{G}$ on $L$,
    then $r(-1)=\pm \unit$.
    \begin{definition}
     An \irrv \repv of $\bar{G}$ with $r(-1)=\unit$ is called a
     \svrepv of $G$ while an \irrv \repv with $r(-1)=-\unit$ is
     called a \spinrepv or \tvrepv of $G$.
    \end{definition}
    \begin{proposition}
    \label{repinduce}
     Let $\irrc{G}$ be the class of all in\idv \irrv
     \rep s of $G$ and $\irrc{G}^s$ be the class of all
     in\idv \svrep s of $G$,
     define $\phi:\irrc{G}\rightarrow \irrc{G}^s, r\mapsto r\circ
     \pi$. Then $\phi$ is a bijection.
    \end{proposition}
    \prf
     One can check: $r\circ \pi$ is a \repv of $\bar{G}$; if
     $r\cong r^\prime$, then $r\circ \pi\cong r^\prime\circ\pi$;
     that $r$ is \irrv implies that $r\circ\pi$ is \irrv and
     $r\circ\pi$ is single-valued. Therefore, $\phi$ is
     well-defined.
     If $r$ and $r^\prime$ are inequivalent, then $r\circ \pi$ and $r^\prime\circ\pi$ are
     two elements in $\irrc{G}^s$, namely $\phi$ is injective. To prove that $\phi$ is a
     surjection, consider any $\tilde{r}\in \irrc{G}^s:\bar{G}\rightarrow L$. Define
     $r:G\rightarrow L, g\mapsto\tilde{r}(s(g))$ where $s(g)$ is any element in $\pi^{-1}(g)$.
     One can check: $r$ is a well-defined map since $\tilde{r}$ is
     single-valued; $r$ is an \irrv \repv of $G$ on $L$,
     accordingly $r\in\irrc{G}$ and lastly, $\phi(r)=\tilde{r}$. So the result follows.\\
    \endprf\\
    This proposition says that all single-valued \rep s of $G$ which are part of inequivalent \irrv
    \rep s of $\bar{G}$ are completely determined by
    the \repv theory of $G$.
  \section{Structure of $\ohd$}
   \subsection{Structure of $\oh$}
   \label {sec.str}
    It follows Proposition \ref{On} that $\oh\cong \ztf\smdp S_4$; hence $|\oh|=384$. In
    point \gpv theory, rotation \subgv of $O_h$ is denoted as $O$;
    on the other hand, $\per{4}\cong \ztt\smdp S_3\cong O$ \cite{s}. We write the
    isomorphism explicitly. The structure of $\ztt \smdp S_3$ is given
    by four generators $\alpha, \beta, \eta, t$ and the relations
    \eqa
    \label{g1}
     \alpha^2=e, \beta^2=e, \alpha\beta=\beta\alpha\\
    \label{g2}
     t^3=e, \eta^2=e, \eta t=t^2\eta\\
    \label{g3}
     t\alpha=\alpha\beta t, t\beta=\alpha t,
     \eta \alpha=\beta \eta
    \ena
    and the isomorphisms are defined to be
    \eqann
     (12)(34)\leftrightarrow \alpha\leftrightarrow diag(-1,-1,1),
     (13)(24)\leftrightarrow \beta\leftrightarrow diag(1,-1,-1)\\
     (234)\leftrightarrow t\leftrightarrow
     \left(
      \begin{array}{lll}
       0&1&0\\
       0&0&1\\
       1&0&0
      \end{array}
     \right),
     (23)\leftrightarrow \eta\leftrightarrow
     \left(
      \begin{array}{lll}
       0&0&-1\\
       0&-1&0\\
       -1&0&0
      \end{array}
     \right)
    \enann
    The structure of $\ztf\smdp S_4$ is given by
    (\ref{g1})(\ref{g2})(\ref{g3}) together with (see Eq.(\ref{genr}))
    \eqa
    \label{o4g1}
     I_i^2=e, I_iI_j=I_jI_i, i,j=1..4, i\not= j\\
    \nn
     \alpha I_1=I_2\alpha, \alpha I_3=I_4\alpha\\
    \nn
     \beta I_1=I_3\beta,\beta I_2=I_4\beta\\
    \label{o4g2}
     t I_1=I_1 t,t I_2=I_4 t,t I_3=I_2 t,t I_4=I_3 t\\
    \nn
     \eta I_1=I_1\eta,\eta I_2=I_3\eta,\eta I_4=I_4\eta
    \ena
    The matrix representations of above generators are given by (see Eq.(\ref{cong}))
    \eqa
    \label{r41}
     (I_i)^j_k=\dlt^j_k(1-2\dlt^j_i), i,j,k=1,2,3,4\\
    \label{r42}
     \alpha\mapsto
      \left(
       \begin{array}{llll}
        0&1&0&0\\
        1&0&0&0\\
        0&0&0&1\\
        0&0&1&0
       \end{array}
      \right)
     \beta\mapsto
      \left(
       \begin{array}{llll}
        0&0&1&0\\
        0&0&0&1\\
        1&0&0&0\\
        0&1&0&0
       \end{array}
      \right)
     t\mapsto
      \left(
       \begin{array}{llll}
        1&0&0&0\\
        0&0&1&0\\
        0&0&0&1\\
        0&1&0&0
       \end{array}
      \right)
     \eta\mapsto
      \left(
       \begin{array}{llll}
        1&0&0&0\\
        0&0&1&0\\
        0&1&0&0\\
        0&0&0&1
       \end{array}
      \right)
    \ena
    In fact, if we introduce
    \[
     \gamma\mapsto
      \left(
       \begin{array}{llll}
        0&0&1&0\\
        0&1&0&0\\
        1&0&0&0\\
        0&0&0&1
       \end{array}
      \right)
    \]
    then the generators of $\oh$ can be reduced to a smaller
    set $\{I_i, \gamma, t|i=1,2,3,4\}$
    whose generating relations are
    (\ref{o4g1})(\ref{o4g2}) together with
    \eqa
    \label{later1}
     \gamma^2=e, t^3=e,(t\gamma)^4=e\\
    \label{later2}
     \gamma I_1=I_3\gamma,\gamma I_2=I_2\gamma,\gamma I_4=I_4\gamma
    \ena
    while $\alpha=(t^2\gamma)^2,\beta=t\gamma t^2\gamma t, \eta=\gamma t\gamma
     t^2\gamma$.\\
     \\
    Applying the general results on \conv \clsfv of $\cgn$
    Eqs.(\ref{syt})(\ref{n1})...(\ref{n4}),
    we give the table of \conv classes of $\oh$ (see Tab. \ref{tab1}).
    \begin{table}[d]
    \caption{Conjugate Classes of $\ztn{4}\smdp\per{4}$}
    \label{tab1}
     \begin{tabbing}
      xxx\=xxxxxxxx\=xxxxxxxxxxxxxxxxx\=xxx\=xxxxx\=xxxxxxxxxx\=xxx\=xxxxxxxx\=xxxxxxxxxxxxxxxxx\=xxx\=xxxxx\=xxx \kill
      No\>SplitNo\>YoungDiagram\>ord\>num\>det \>No\>SplitNo\>YoungDiagram\>ord\>num\>det\\
      1\>1-1\>\ybxa{0}{0}{0}{0}\>1\>1\>1\>2\>1-2\>\ybxa{0}{0}{0}{10}\>2\>4\>-1\\
      3\>1-3\>\ybxa{0}{0}{10}{10}\>2\>6\>1\>4\>1-4\>\ybxa{0}{10}{10}{10}\>2\>4\>-1\\
      5\>1-5\>\ybxa{10}{10}{10}{10}\>2\>1\>1\\
      6\>2-1\>\ybxb{0}{0}{0}{0}\>2\>12\>-1\>7\>2-2\>\ybxb{0}{0}{0}{10}\>2\>24\>1\\
      8\>2-3\>\ybxb{10}{0}{0}{0}\>4\>12\>1\>9\>2-4\>\ybxb{0}{0}{10}{10}\>2\>12\>-1\\
      10\>2-5\>\ybxb{10}{0}{0}{10}\>4\>24\>-1\>11\>2-6\>\ybxb{10}{0}{10}{10}\>4\>12\>1\\
      12\>3-1\>\ybxc{0}{0}{0}{0}\>2\>12\>1\>13\>3-2\>\ybxc{0}{10}{0}{0}\>4\>24\>-1\\
      14\>3-3\>\ybxc{10}{10}{0}{0}\>4\>12\>1\\
      15\>4-1\>\ybxd{0}{0}{0}{0}\>3\>32\>1\>16\>4-2\>\ybxd{0}{0}{0}{10}\>6\>32\>-1\\
      17\>4-3\>\ybxd{10}{0}{0}{0}\>6\>32\>-1\>18\>4-4\>\ybxd{10}{0}{0}{10}\>6\>32\>1 \\
      19\>5-1\>\ybxe{0}{0}{0}{0}\>4\>48\>-1\>20\>5-2\>\ybxe{10}{0}{0}{0}\>8\>48\>1  \\
     \end{tabbing}
     ``SplitNo" reflects the relation between the
     classes of $\zt\wr\per{4}$ and those of $\per{4}$. ``ord" means order of each class.
     ``num" is the number of elements in each class. ``det" is the
     signature of each class. See
     Eqs.(\ref{n1})...(\ref{n4}).
    \end{table}
   \subsection{Construction of $\ohd$}
   \label{ohdstr}
    We will denote $s(g)$ still as $g$ for all $g\in G$.
    $\ohd$ is generated by the equations below.
    \begin{proposition}
    \label{gen}
     \eqa
     \label{o4bg1}
      I_i^2=-1, I_iI_j=-I_jI_i, i,j=1..4, i\not= j\\
     \label{o4bg2}
      \gamma^2=-1, t^3=-1, (t\gamma)^4=-1\\
     \label{o4bg3}
      {\underline{\gamma I_1=-I_3\gamma}}, \gamma I_2=-I_2\gamma, \gamma I_4=-I_4\gamma\\
     \label{o4bg4}
      tI_1=I_1t, {\underline{tI_2=I_4t}},{\underline{tI_3=I_2t}},
      {\underline{tI_4=I_3t}}
     \ena
    \end{proposition}
    \prf
     First
     Eqs.(\ref{o4bg1})...(\ref{o4bg4}) are
     valid. In fact, the standard orthogonal bases in $E^4$ satisfy Clifford
     relations
     $e_ie_j+e_je_i=-2\delta_{ij}$ which is
     equivalent to (\ref{o4bg1}); therefore, one can take
     $I_i=e_i$. Following lemma \ref{reflection}, we set
     $\gamma={1\over{\sqrt{2}}}(e_3-e_1)$ and check that
     (\ref{o4bg3}) is satisfied. Let $t={1\over 2}(1-e_2e_3+e_2e_4-e_3e_4)$ which is the product of
     ${1\over{\sqrt{2}}}(e_2-e_3)$ and
     ${1\over{\sqrt{2}}}(e_4-e_2)$, and (\ref{o4bg4}) can be
     verified. Finally, one can check that (\ref{o4bg2}) is also
     satisfied.\\
     \\
     Second, notice that above equations are just Eqs.(\ref{o4g1})(\ref{o4g2})(\ref{later1})(\ref{later2}),
     which generate $\oh$, twisted by a $\zt$ factor set.
     So due to the validity of the above equations, $\forall g\in \ohd
     $, either $g$ or $-g$ will be generated. But $-1$ can be
     generated. Therefore, the above equation set generations
     $\ohd$.\\
    \endprf
    We can give another proof of this result by proposition \ref{rep}.
    In fact, we introduce $\gamma$-matrices in $E^4$ as
    \[
     \gamma_i=\left(
      \begin{array}{cc}
       \zrt&i\sigma_i\\i\sigma_i&\zrt
      \end{array}\right),i=1,2,3;
     \gamma_4=\left(
      \begin{array}{cc}
       \zrt&-\unt\\ \unt&\zrt
      \end{array}\right)
    \]
    in which $\sigma_i$ stand for three Pauli matrices
    \[
     \sigma_1=\left(
      \begin{array}{cc}
       0&1\\1&0
      \end{array}\right),
     \sigma_2=\left(
      \begin{array}{cc}
       0&i\\-i&0
      \end{array}\right),
     \sigma_3=\left(
      \begin{array}{cc}
       1&0\\0&-1
      \end{array}\right)
    \]
    Note that $\sigma_2$ in our convention is different from the
    usual definition in physics.
    $\gamma_i (i=1..4)$ satisfy Clifford relations $\gamma_i\gamma_j
    +\gamma_j\gamma_i=-2\delta_{ij}\unf$ and $
     \gamma_i^\dag=-\gamma_i, \gamma_i \gamma_i^\dag=\unf,
     det(\gamma_i)=1$.\\
     \\
    We use $S(g)$ as the image of $s(g)$ in $M_2(\quaternion)$. Let
    \eq
    \label{L1}
     S(I_i)=\gamma_i,
     S(\gamma)={i\over\sqrt{2}}\cdot
      \left(
       \begin{array}{cccc}
        0&0&1&-1\\0&0&-1&-1\\1&-1&0&0\\-1&-1&0&0
       \end{array}
      \right),
     S(t)={e^{i\frac{7\pi}{4}}\over\sqrt{2}}\cdot
      \left(
       \begin{array}{cccc}
        1&-i&0&0\\1&i&0&0\\0&0&i&1\\0&0&-i&1
       \end{array}
      \right)
    \en
    then one can check that these matrices give correct images under $\widetilde{Ad}$ and satisfy
    the corresponding relations in (\ref{o4bg1})...(\ref{o4bg4}).
    It should be noticed that the $\widetilde{Ad}$-map condition can
    fix these matrices up to a non-vanishing scalar and that by
    using lemma \ref{su4}, the scalar can be fixed up to a $Z_4$
    uncertainty, namely if one searches out a $S(g)$ then
    $iS(g),-S(g),-iS(g)$ will also work. One can figure out
    two of them by calculating the projections on the basis of
    $Cl(E^4)$ and ruling out those whose projections are pure
    imagine. \\
    \\
    We point out that the generating relations
    in proposition \ref{gen} are not unique, due to the canonical
    automorphism of $Cl(E^4)$. In fact from the second proof of
    this proposition, we have notified that at last there is still
    a $\zt$ uncertainty. Consequently, we can change the cross-section $s$ to
    another one $s^\prime$ by a ``local" $\zt$ transformation and the
    underlined equations in
    Eqs.(\ref{o4bg1})...(\ref{o4bg4}) may gain or lose some
    $(-1)$-factors accordingly. Anyway, they are equivalent to the
    former ones.\\
    \\
    To classify the elements in $\ohd$, Lemma \ref{conjugat} will
    enable us to use the same symbols for the \conv classes of
    $\oh$ and to use a ``$\pr$" for those splitting classes. Except for classes $1,8,14,15,20$
    which split into two classes for each, any other class in $\oh$ is lifted to one class. Therefore,
    there are totally  $25$ classes in $\ohd$ (see table \ref{tab2}).
    \begin{table}[d]
    \renewcommand{\arraystretch}{1.4}
    \caption{Conjugate Classes of $\ohd$}
    \label{tab2}
    \scriptsize
    \begin{tabular}{|l|r|r|r|r|r|r|r|r|r|r|r|r|r|r|r|r|r|r|r|r|r|r|r|r|r|}\hline\hline
     No.&1&$1^\pr$&2&3&4&5&6&7&8&$8^\pr$&9&10&11&12&13&14&$14^\pr$&15&$15^\pr$&16&17&18&19&20&$20^\pr$\\ \hline
     num.&1&1&8&12&8&2&24&48&12&12&24&48&24&24&48&12&12&32&32&64&64&64&96&48&48\\ \hline
     ord.&1&2&4&4&2&2&4&4&8&8&2&8&8&4&8&4&4&6&3&12&6&6&4&8&8\\
     \hline\hline
    \end{tabular}
    \normalsize
    \\
    \\
    The labels of classes are descended from those of $\oh$ with
    ``$\pr$" for those classes split when lifted into $\ohd$.
    \end{table}
  \section{\Rep s of $\ohd$}
  \label{repofohd}
   \subsection{Single-valued \rep s of $\oh$}
    Due to Theorem \ref{repinduce}, there are totally 20 in\idv \svrep s
    of $\oh$ corresponding to the 20 in\idv \irrv \rep s of $\oh$; the \repthv of $\oh$
    can be systematically solved by applying \ltgpv method
    (Proposition \ref{rfm}).\\

    \begin{table}[d]
    \renewcommand{\arraystretch}{1.4}
    \caption{Character Table of $\ztf$}
    \label{tab3}
    \footnotesize{
     \begin{tabular}{|l|c|c|c|c|c|c|c|c|c|c|c|c|c|c|c|c|}\hline
      $\ztfb$&$[e]$&$[I_1]$&$[I_2]$&$[I_3]$&$[I_4]$
      &$[I_{12}]$&$[I_{13}]$&$[I_{14}]$&$[I_{23}]$&$[I_{24}]$&$[I_{34}]$&$[I_{234}]$&$[I_{134}]$&$[I_{124}]$&$[I_{123}]$
      &$[I_{1234}]$\\ \hline
      $\chi_{0000}$&1&1&1&1&1&1&1&1&1&1&1&1&1&1&1&1\\ \hline
      $\chi_{0001}$&1&1&1&1&-1&1&1&-1&1&-1&-1&-1&-1&-1&1&-1\\ \hline
      $\chi_{0010}$&1&1&1&-1&1&1&-1&1&-1&1&-1&-1&-1&1&-1&-1\\ \hline
      $\chi_{0100}$&1&1&-1&1&1&-1&1&1&-1&-1&1&-1&1&-1&-1&-1\\ \hline
      $\chi_{1000}$&1&-1&1&1&1&-1&-1&-1&1&1&1&1&-1&-1&-1&-1\\ \hline
      $\chi_{0011}$&1&1&1&-1&-1&1&-1&-1&-1&-1&1&1&1&-1&-1&1\\ \hline
      $\chi_{0101}$&1&1&-1&1&-1&-1&1&-1&-1&1&-1&1&-1&1&-1&1\\ \hline
      $\chi_{1001}$&1&-1&1&1&-1&-1&-1&1&1&-1&-1&-1&1&1&-1&1\\ \hline
      $\chi_{0110}$&1&1&-1&-1&1&-1&-1&1&1&-1&-1&1&-1&-1&1&1\\ \hline
      $\chi_{1010}$&1&-1&1&-1&1&-1&1&-1&-1&1&-1&-1&1&-1&1&1\\ \hline
      $\chi_{1100}$&1&-1&-1&1&1&1&-1&-1&-1&-1&1&-1&-1&1&1&1\\ \hline
      $\chi_{1110}$&1&-1&-1&-1&1&1&1&-1&1&-1&-1&1&1&1&-1&-1\\ \hline
      $\chi_{1101}$&1&-1&-1&1&-1&1&-1&1&-1&1&-1&1&1&-1&1&-1\\ \hline
      $\chi_{1011}$&1&-1&1&-1&-1&-1&1&1&-1&-1&1&1&-1&1&1&-1\\ \hline
      $\chi_{0111}$&1&1&-1&-1&-1&-1&-1&-1&1&1&1&-1&1&1&1&-1\\ \hline
      $\chi_{1111}$&1&-1&-1&-1&-1&1&1&1&1&1&1&-1&-1&-1&-1&1\\ \hline
     \end{tabular}}
    \normalsize
    \\
    \\
    $I_{i_1i_2...I_a}:=I_{i_1}\cdot I_{i_2}\cdot ...\cdot I_{i_a}$. Irreducible characters are labeled as $\chi_{s_1s_2s_3s_4}, s_i\in
    \intg/2\intg$ (see Eq.(\ref{zrp})).
    \end{table}
    All in\idv \irrv characters of $\ztf$ are listed in Table \ref{tab3}.
    Following Theorem \ref{ron}, $\Pi(\ztf)$ are partitioned into orbits with index set defined in
    a physical convention $\indxst:=\{S,P,V,A,T\}$.
    \eqann
     \Pi_S=\{\pi_{0000}\}, F_S\cong S_4;\Pi_P=\{\pi_{1111}\}, F_P\cong S_4;\\
     \Pi_V=\{\pi_{0001},\pi_{0010},\pi_{0100},\pi_{1000}\}, F_V\cong S_3;\\
     \Pi_A=\{\pi_{1110},\pi_{1101},\pi_{1011},\pi_{0111}\}, F_A\cong S_3;\\
     \Pi_T=\{\pi_{0011},\pi_{0101},\pi_{1001},\pi_{0110},\pi_{1010},\pi_{1100}\}, F_T\cong \ztt
    \enann
    We will use $[\lam]$ instead of $(\nu)$ to denote Young
    diagrams where $[\lam]=[\lam_1\lam_2...\lam_n],
    \lam_k=\sum_{i=k}^n{\nu_i}$.\\
    \\
    \orb S\\
    \label{orb1}
     All in\idv \irrv \rep s of $S_4$ is labeled by
     $[4],[31],[2^2],[21^2],[1^4]$; accordingly,
     \[
      \Pi_S\cdot([4],[31],[2^2],[21^2],[1^4])
     \]
     provide two one-\diml, one two-\dimlv and two three-\dimlv
     \rep s. As for \repv matrices, all $I_i,i=1..4$ are mapped to identity, while $\alpha,
     \beta,t,\eta$ take the same matrix form as they have in $S_4$, i.e.
     $\Pi_S\cdot[4]:
       I_i,\alpha,\beta,t,\eta\rightarrow 1$;
     $\Pi_S\cdot[1^4]:
       I_i,\alpha,\beta,t\rightarrow 1,\eta\rightarrow -1$;
     \[
      \Pi_S\cdot[2^2]:
       I_i,\alpha,\beta\rightarrow\unt,
       t\rightarrow
        \left(
         \begin{array}{cc}
          e^{i\frac{2\pi}{3}}&0\\0&e^{i\frac{4\pi}{3}}
         \end{array}
        \right),
       \eta\rightarrow
        \left(
         \begin{array}{cc}
          0&1\\1&0
         \end{array}
        \right);
      \]
      \[
      \Pi_S\cdot[31]:
       I_i\rightarrow\unth,
       \alpha\rightarrow
        \left(
         \begin{array}{ccc}
          -1&0&0\\0&1&0\\0&0&-1
         \end{array}
        \right),
       \beta\rightarrow
        \left(
         \begin{array}{ccc}
          -1&0&0\\0&-1&0\\0&0&1
         \end{array}
        \right),
       t\rightarrow
        \left(
         \begin{array}{ccc}
          0&0&1\\1&0&0\\0&1&0
         \end{array}
        \right),
       \eta\rightarrow
        \left(
         \begin{array}{ccc}
          1&0&0\\0&0&1\\0&1&0
         \end{array}
        \right);
      \]
    $\Pi_S\cdot[21^2]$: $I_i, \alpha, \beta, t$ take the same form
    of $\Pi_S\cdot[31]$ and $\eta$ gains a minus sign compared to
    $\Pi_S\cdot[31]$.\\
    \\
  \orb P\\
   \[
    \Pi_P\cdot([4],[31],[2^2],[21^2],[1^4])
   \]
   The only difference from \orb S is that $I_i$ are mapped to
   $-\unit$.\\
   \\
  \orb V\\
  \label{orb3}
   All in\idv \irrv \rep s of $S_3$ can be written as $[3],[21],[1^3]$
   and it has no difficulty,
   using our generating relations, to check
   \[
    \alpha\pi_{1000}\alpha^{-1}=\pi_{0100},
    \eta\pi_{0100}\eta^{-1}=\pi_{0010},
    \alpha\pi_{0010}\alpha^{-1}=\pi_{0001}
   \]
   Hence, this orbit gives two four-\dimlv \rep s and one eight-\dimlv
   \rep.
   \[
    \Pi_V\cdot ([3],[21],[1^3])=(e,\alpha,\beta\eta,\alpha\beta\eta)\cdot\pi_{1000}\cdot ([3],[21],[1^3])
   \]
   The \repv matrices of $\Pi_V\cdot [3]$ are coincident with those in Eqs.(\ref{r41})(\ref{r42}).
   \Repv matrices of $\Pi_V\cdot [1^3]$ are the same as those in $\Pi_V\cdot [3]$,
   except that $\eta$ picking on a minus sign. \\
   \\
   $\Pi_V\cdot [21]:$
    \[
     e_i\rightarrow
      \left(
       \begin{array}{cc}
        \Pi_V\cdot [3](e_i)&\zrf\\
        \zrf&\Pi_V\cdot [3](e_i)
       \end{array}
      \right),
    \]
    \eqann
     \alpha\rightarrow
      \left(
       \begin{array}{cc}
        \Pi_V\cdot [3](\alpha)&\zrf\\
        \zrf&\Pi_V\cdot [3](\alpha)
       \end{array}
      \right),
     \beta\rightarrow
      \left(
       \begin{array}{cc}
        \zrf&\Pi_V\cdot [3](\beta)\\
        \Pi_V\cdot [3](\beta)&\zrf
       \end{array}
      \right),\\
     t\rightarrow
      \left(
       \begin{array}{cc}
        e^{i\frac{2\pi}{3}}\cdot\Pi_V\cdot [3](t)&\zrf\\
        \zrf&e^{i\frac{4\pi}{3}}\cdot\Pi_V\cdot [3](t)
        \end{array}
      \right),
     \eta\rightarrow
      \left(
       \begin{array}{cc}
        \zrf&\Pi_V\cdot [3](\eta)\\
        \Pi_V\cdot [3](\eta)&\zrf
       \end{array}
      \right)
    \enann
  \orb A
  \label{orb4}\\
   Similar to \orb V, there are two four-\dimlv \rep s and one eight-\dimlv
   \rep.
   \[
    \Pi_A\cdot ([3],[21],[1^3])=(e,\alpha,\beta\eta,\alpha\beta\eta)\cdot\pi_{0111}\cdot([3],[21],[1^3])
   \]
   while the \repv matrices for $I_i$ pick on a minus sign, without changing the
   others.\\
   \\
  \orb T\\
   The \stbsbv $F_T$ leaving $\pi_{0110}$ invariant is
   $\{e,\eta,\alpha\beta,\alpha\beta\eta\}$ with
   four one-dimensional \irrv \rep s, denoted by
   $\pi_{(a,b)},a,b=0,1$. Therefore, there are four six-\dimlv \rep s given by
   this orbit. Notice that
   \eqann
    \alpha\pi_{0110}\alpha^{-1}=\pi_{1001},
    t\pi_{1001}t^{-1}=\pi_{1010},
    t\pi_{1010}t^{-1}=\pi_{1100},\\
    \alpha\pi_{1010}\alpha^{-1}=\pi_{0101},
    \beta\pi_{1100}\beta^{-1}=\pi_{0011},
   \enann
   the four \rep s can be labeled as
   \[
    \Pi_T\cdot(\pi_{00},\pi_{01},\pi_{10},\pi_{11})=(e,\alpha,\alpha\beta t,\beta t^2,\beta
    t,t^2)\cdot \pi_{0110}
    \cdot(\pi_{00}+\pi_{01}+\pi_{10}+\pi_{11})
   \]
   Then we enumerate the matrices for the four \rep s.
   \[
    \Pi_T \cdot\pi_{00}:
     I_1\rightarrow diag(1,-1,-1,-1,1,1),I_2\rightarrow diag(-1,1,1,-1,-1,1),
   \]
   \[
    I_3\rightarrow diag(-1,1,-1,1,1,-1),I_4\rightarrow diag(1,-1,1,1,-1,-1),
   \]
   \[
    \alpha\rightarrow
      \left(
       \begin{array}{cccccc}
        0&1&0&0&0&0\\
        1&0&0&0&0&0\\
        0&0&0&0&1&0\\
        0&0&0&1&0&0\\
        0&0&1&0&0&0\\
        0&0&0&0&0&1
       \end{array}
      \right)
     \beta\rightarrow
      \left(
       \begin{array}{cccccc}
        0&1&0&0&0&0\\
        1&0&0&0&0&0\\
        0&0&1&0&0&0\\
        0&0&0&0&0&1\\
        0&0&0&0&1&0\\
        0&0&0&1&0&0
       \end{array}
      \right)
    \]
    \[
     t\rightarrow
      \left(
       \begin{array}{cccccc}
        0&0&0&0&0&1\\
        0&0&0&1&0&0\\
        0&1&0&0&0&0\\
        0&0&1&0&0&0\\
        1&0&0&0&0&0\\
        0&0&0&0&1&0
       \end{array}
      \right)
     \eta\rightarrow
      \left(
       \begin{array}{cccccc}
        1&0&0&0&0&0\\
        0&1&0&0&0&0\\
        0&0&0&1&0&0\\
        0&0&1&0&0&0\\
        0&0&0&0&0&1\\
        0&0&0&0&1&0
       \end{array}
      \right)
   \]
   \[
    \Pi_T \cdot\pi_{01}:
     I_1\rightarrow \Pi_2 \cdot\pi_{00}(I_1),
     I_2\rightarrow \Pi_T \cdot\pi_{00}(I_2),
     I_3\rightarrow \Pi_T \cdot\pi_{00}(I_3),
     I_4\rightarrow \Pi_T \cdot\pi_{00}(I_4)
   \]
   \[
     \alpha\rightarrow
      \left(
       \begin{array}{cccccc}
        0&1&0&0&0&0\\
        1&0&0&0&0&0\\
        0&0&0&0&1&0\\
        0&0&0&-1&0&0\\
        0&0&1&0&0&0\\
        0&0&0&0&0&-1
       \end{array}
      \right)
     \beta\rightarrow
      \left(
       \begin{array}{cccccc}
        0&-1&0&0&0&0\\
        -1&0&0&0&0&0\\
        0&0&-1&0&0&0\\
        0&0&0&0&0&1\\
        0&0&0&0&-1&0\\
        0&0&0&1&0&0
       \end{array}
      \right)
   \]
   \[
     t\rightarrow
      \left(
       \begin{array}{cccccc}
        0&0&0&0&0&1\\
        0&0&0&1&0&0\\
        0&1&0&0&0&0\\
        0&0&1&0&0&0\\
        -1&0&0&0&0&0\\
        0&0&0&0&-1&0
       \end{array}
      \right)
     \eta\rightarrow
      \left(
       \begin{array}{cccccc}
        1&0&0&0&0&0\\
        0&-1&0&0&0&0\\
        0&0&0&-1&0&0\\
        0&0&-1&0&0&0\\
        0&0&0&0&0&-1\\
        0&0&0&0&-1&0
       \end{array}
      \right)
    \]
    \[
    \Pi_T \cdot\pi_{10}:
     I_1\rightarrow \Pi_T \cdot\pi_{00}(I_1),
     I_2\rightarrow \Pi_T \cdot\pi_{00}(I_2),
     I_3\rightarrow \Pi_T \cdot\pi_{00}(I_3),
     I_4\rightarrow \Pi_T \cdot\pi_{00}(I_4)
    \]
    \[
     \alpha\rightarrow \Pi_T \cdot\pi_{00}(\alpha),
     \beta\rightarrow \Pi_T \cdot\pi_{00}(\beta),
     t\rightarrow \Pi_T \cdot\pi_{00}(t),
     \eta\rightarrow (-1)\cdot\Pi_T \cdot\pi_{00}(\eta)
    \]
    \[
    \Pi_T \cdot\pi_{11}:
     I_1\rightarrow \Pi_T \cdot\pi_{01}(I_1),
     I_2\rightarrow \Pi_T \cdot\pi_{01}(I_2),
     I_3\rightarrow \Pi_T \cdot\pi_{01}(I_3),
     I_4\rightarrow \Pi_T \cdot\pi_{01}(I_4)
    \]
    \[
     \alpha\rightarrow \Pi_T \cdot\pi_{01}(\alpha),
     \beta\rightarrow \Pi_T \cdot\pi_{01}(\beta),
     t\rightarrow \Pi_T \cdot\pi_{01}(t),
     \eta\rightarrow (-1)\cdot\Pi_T \cdot\pi_{01}(\eta)
    \]
  Here we find all 20 inequivalent \irrv \rep s corresponding
  to the 20 \conv classes of $\oh$, which satisfy Burside formula
  \[
   2\times(1^2+1^2+2^2+3^2+3^2)+2\times(4^2+4^2+8^2)+4\times 6^2
   =384
  \]
  Following proposition \ref{repinduce}, we have found all of the
  single-valued \rep s of $\oh$.
 \subsection{Spinor \rep s of $\oh$}
  Notice the following facts that $\ov{\ztf}\lhd\ohd$, $\ohd/\ov{\ztf}\cong
  \per{4}$ and Eqs.(\ref{L1}) generate a \spinrepv of
  $\oh$ which is denoted still as $S$; what's more, its restriction to $\ov{\ztf}$
  is also a \tvrepv of $\ztf$. These facts ensure two conditions in Theorem
  \ref{clifford}. To apply Theorem \ref{clifford} to deduce \spinrep s of $\oh$, we develop a
  calculation method. The matrices of a \spinrepv of $\oh$ for $I_i,\gamma,t$, denoted
  as $\tld{S}(I_i),\tld{S}(\gamma),\tld{S}(t)$, can be decomposed as
  \eqann
   \tld{S}(I_i)=S(I_i)\ot\unit,i=1,3,4;\tld{S}(I_2)=-S(I_2)\ot\unit;\\
   \tld{S}(\gamma)=\Gamma\ot \tld{\gamma};
   \tld{S}(t)=T\ot\tilde{t}
  \enann
  where $\Gamma,T$ and $S(I_i)$ act on the same module,
  $\tilde{\gamma},\tilde{t}$ have \dfn{the same
  texture (zero matrix elements) of the \repv matrices of five in\idv \irrv \rep s of $S_4$} (
  the minus added before $S(I_2)$ is for a physical convention).
  There are five \spinrep s of dimension 4,4,8,12 and 12 respectively and
  the second half of Burside formula is satisfied.
  \[
   4^2+4^2+8^2+12^2+12^2=384
  \]
  Corresponding the generating equations
  (\ref{o4bg1})...(\ref{o4bg4}), there are
  a system of matrix equations.
  \eq
  \label{4deq1}
   \tld{S}(\gamma)^2=\tld{S}(t)^3=-\unit,(\tld{S}(\gamma)\tld{S}(t))^4=-\unit
  \en
  \eq
  \label{4deq2}
   \tld{S}(\gamma)\tld{S}(I_2)=-\tld{S}(I_2)\tld{S}(\gamma),
   \tld{S}(\gamma)\tld{S}(I_4)=-\tld{S}(I_4)\tld{S}(\gamma),
   \tld{S}(\gamma)\tld{S}(I_1)=-\tld{S}(I_3)\tld{S}(\gamma),
  \en
  \eq
  \label{4deq3}
   \tld{S}(t)\tld{S}(I_1)=\tld{S}(I_1)\tld{S}(t),
   \tld{S}(t)\tld{S}(I_2)=-\tld{S}(I_4)\tld{S}(t),
   \tld{S}(t)\tld{S}(I_3)=-\tld{S}(I_2)\tld{S}(t),
   \tld{S}(t)\tld{S}(I_4)=\tld{S}(I_3)\tld{S}(t),
  \en
  plus a unitary condition
  \eq
  \label{4deq4}
   \tld{S}(\gamma)^{\dagger}\tld{S}(\gamma)=\unit,
   \tld{S}(t)^{\dagger}\tld{S}(t)=\unit
  \en
  Note that we add a minus sign to the second and the third equations in
  Eq.(\ref{4deq3}) compared with Eq.(\ref{o4bg4}) according to the same physics convention, though they
  are completely equivalent.\\
  \\
  Solving Eqs.(\ref{4deq1})...(\ref{4deq4}) for four-\dimlv case gives two solutions
  \[
   \b{4}_{+}:
    \tld{S}(\gamma)=\Gamma\cdot \tilde{\gamma}_{+},
    \Gamma={1\over\sqrt{2}}\cdot
     \left(
      \begin{array}{cccc}
       0&0&1&-1\\0&0&-1&-1\\1&-1&0&0\\-1&-1&0&0
      \end{array}
     \right),
    \tilde{\gamma}_{+}=i
  \]
  \[
    \tld{S}(t)=T\cdot\tilde{t},
    T={1\over\sqrt{2}}\cdot
     \left(
      \begin{array}{cccc}
       1&-i&0&0\\1&i&0&0\\0&0&i&1\\0&0&-i&1
      \end{array}
     \right),
    \tilde{t}=e^{i\frac{7\pi}{4}}
  \]
  \[
   \b{4}_{-}:
    \tld{S}(\gamma)=\Gamma\cdot \tilde{\gamma}_{-},
    \tilde{\gamma}_{-}=-i,
    \tld{S}(t)=T\cdot\tilde{t}
  \]
  Note that $\b{4}_{+}$ is just the \repv $S$ with $\tld{S}(I_2)=-S(I_2)$.\\
  \\
  As for eight-\dimlv case we can suppose
  \[
   \tld{\gamma}=
    \left(
     \begin{array}{cc}
      0&\tilde{c}\\\tilde{d}&0
     \end{array}
    \right),
   \tld{t}=
    \left(
     \begin{array}{cc}
      \tilde{a}&0\\0&\tilde{d}
     \end{array}
    \right)
  \]
   The solution $\b{8}$ is given by
  \[
   \tld{c}=e^{i\frac{\pi}{3}},\tld{d}=e^{i\frac{2\pi}{3}},
   \tld{a}=e^{i\frac{5\pi}{12}},\tld{d}=e^{i\frac{13\pi}{12}}
  \]
  Finally, we set for the twelve-\dimlv case
  \[
   \tld{\gamma}=
    \left(
     \begin{array}{ccc}
      0&0&\tld{z}\\0&\tld{y}&0\\ \tld{x}&0&0
     \end{array}
    \right),
   \tld{t}=
    \left(
     \begin{array}{ccc}
      0&0&\tld{n}\\ \tld{l}&0&0\\ 0&\tld{m}&0
     \end{array}
    \right)
  \]
  Such that
  \eqann
   \b{12}_+:
    \tld{x}=1,\tld{y}=i,\tld{z}=-1,
    \tld{l}=1,\tld{m}=1,\tld{n}=e^{i\frac{5\pi}{4}}\\
   \b{12}_-:
    \tld{x}=1,\tld{y}=-i,\tld{z}=-1,
    \tld{l}=1,\tld{m}=-1,\tld{n}=e^{i\frac{\pi}{4}}
  \enann
  So far, we obtain all in\idv \irrv \rep s of $\ohd$ and we summarize our results in Table \ref{tabch}.
  \begin{table}[hd]
  \caption{Character Table of $\ohd$}
  \label{tabch}
  \footnotesize
  \tabcolsep .7pt
  \begin{tabular}{|l||c|c|c|c|c|c|c|c|c|c|c|c|c|c|c|c|c|c|c|c||c|c|c|c|c|}\hline
   &\multicolumn{5}{c|}{$\Pi_S$}&\multicolumn{5}{c|}{$\Pi_P$}&
   \multicolumn{3}{c|}{$\Pi_V$}&\multicolumn{3}{c|}{$\Pi_A$}&\multicolumn{4}{c|}{$\Pi_T$}&\multicolumn{5}{c|}{spinor
   rep.}\\ \cline{2-26}
   &$[4]$&$[1^4]$&$[2^2]$&$[31]$&$[21^2]$&$[4]$&$[1^4]$&$[2^2]$&$[31]$&$[21^2]$
   &$[3]$&$[1^3]$&$[21]$&$[3]$&$[1^3]$&$[21]$&$\pi_{00}$&$\pi_{01}$&$\pi_{10}$&$\pi_{11}$
      &$\b{4}_+$&$\b{4}_-$&$\b{8}$&$\b{12}_+$&$\b{12}_-$\\ \hline
   &$\chi^{(1)}_{1}$
   &$\chi^{(1)}_{3}$&$\chi^{(2)}_{1}$&$\chi^{(3)}_{1}$&
   $\chi^{(3)}_{3}$&$\chi^{(1)}_{2}$&$\chi^{(1)}_{4}$&$\chi^{(2)}_{2}$&$\chi^{(3)}_{2}$&$\chi^{(3)}_{4}$
   &$\chi^{(4)}_{1}$&$\chi^{(4)}_{3}$&$\chi^{(8)}_{1}$&$\chi^{(4)}_{4}$&$\chi^{(4)}_{2}$&$\chi^{(8)}_{2}$
   &$\chi^{(6)}_{3}$&$\chi^{(6)}_{2}$&$\chi^{(6)}_{4}$&$\chi^{(6)}_{1}$
      &$\chi^{(\underline{4})}_{1}$&$\chi^{(\underline{4})}_{2}$&$\chi^{(\underline{8})}$&$\chi^{(\underline{12})}_{1}$
      &$\chi^{(\underline{12})}_{2}$\\ \hline  \hline
                 1&1&1&2&3&3&1&1&2&3&3&4&4&8&4&4&8&6&6&6&6&4&4&8&12&12\\ \hline
                 $1^\pr$&1&1&2&3&3&1&1&2&3&3&4&4&8&4&4&8&6&6&6&6&-4&-4&-8&-12&-12\\ \hline
                 2&1&1&2&3&3&-1&-1&-2&-3&-3&2&2&4&-2&-2&-4&0&0&0&0&0&0&0&0&0\\ \hline
                 3&1&1&2&3&3&1&1&2&3&3&0&0&0&0&0&0&-2&-2&-2&-2&0&0&0&0&0\\ \hline
                 4&1&1&2&3&3&-1&-1&-2&-3&-3&-2&-2&-4&2&2&4&0&0&0&0&0&0&0&0&0\\ \hline
                 5&1&1&2&3&3&1&1&2&3&3&-4&-4&-8&-4&-4&-8&6&6&6&6&0&0&0&0&0\\ \hline
                 6&1&-1&0&1&-1&1&-1&0&1&-1&2&-2&0&2&-2&0&2&0&-2&0&0&0&0&0&0\\ \hline
                 7&1&-1&0&1&-1&-1&1&0&-1&1&0&0&0&0&0&0&0&2&0&-2&0&0&0&0&0\\ \hline
                 8&1&-1&0&1&-1&-1&1&0&-1&1&2&-2&0&-2&2&0&0&-2&0&2&$-2\sqrt{2}$&$2\sqrt{2}$&0&$-2\sqrt{2}$&$2\sqrt{2}$\\ \hline
         $8^\pr$&1&-1&0&1&-1&-1&1&0&-1&1&2&-2&0&-2&2&0&0&-2&0&2&$2\sqrt{2}$&$-2\sqrt{2}$&0&$2\sqrt{2}$&$-2\sqrt{2}$\\ \hline
                 9&1&-1&0&1&-1&1&-1&0&1&-1&-2&2&0&-2&2&0&2&0&-2&0&0&0&0&0&0\\ \hline
                10&1&-1&0&1&-1&1&-1&0&1&-1&0&0&0&0&0&0&-2&0&2&0&0&0&0&0&0\\ \hline
                11&1&-1&0&1&-1&-1&1&0&-1&1&-2&2&0&2&-2&0&0&-2&0&2&0&0&0&0&0\\ \hline
                12&1&1&2&-1&-1&1&1&2&-1&-1&0&0&0&0&0&0&2&-2&2&-2&0&0&0&0&0\\ \hline
                13&1&1&2&-1&-1&-1&-1&-2&1&1&0&0&0&0&0&0&0&0&0&0&0&0&0&0&0\\ \hline
                14&1&1&2&-1&-1&1&1&2&-1&-1&0&0&0&0&0&0&-2&2&-2&2&2&2&4&-2&-2\\ \hline
   $14^\pr$&1&1&2&-1&-1&1&1&2&-1&-1&0&0&0&0&0&0&-2&2&-2&2&-2&-2&-4&2&2\\ \hline
                15&1&1&-1&0&0&1&1&-1&0&0&1&1&-1&1&1&-1&0&0&0&0&2&2&-2&0&0\\ \hline
   $15^\pr$&1&1&-1&0&0&1&1&-1&0&0&1&1&-1&1&1&-1&0&0&0&0&-2&-2&2&0&0\\ \hline
                16&1&1&-1&0&0&-1&-1&1&0&0&-1&-1&1&1&1&-1&0&0&0&0&0&0&0&0&0\\ \hline
                17&1&1&-1&0&0&-1&-1&1&0&0&1&1&-1&-1&-1&1&0&0&0&0&0&0&0&0&0\\ \hline
                18&1&1&-1&0&0&1&1&-1&0&0&-1&-1&1&-1&-1&1&0&0&0&0&0&0&0&0&0\\ \hline
                19&1&-1&0&-1&1&1&-1&0&-1&1&0&0&0&0&0&0&0&0&0&0&0&0&0&0&0\\ \hline
                20&1&-1&0&-1&1&-1&1&0&1&-1&0&0&0&0&0&0&0&0&0&0&$\sqrt{2}$&$-\sqrt{2}$&0&$-\sqrt{2}$&$\sqrt{2}$\\ \hline
   $20^\pr$&1&-1&0&-1&1&-1&1&0&1&-1&0&0&0&0&0&0&0&0&0&0&$-\sqrt{2}$&$\sqrt{2}$&0&$\sqrt{2}$&$-\sqrt{2}$\\ \hline
  \end{tabular}
  \normalsize
  \\
  \\
   Character is labeled by a superscript showing its \dimv where an underline shows a \spinrepv
   and a subscript distinguishing different \rep s with same
   \dim.
  \end{table}
  \section{Structure of $\overline{SO_4}$}\label{so4case}
   Specify $n=4$ in Eq.(\ref{son}) then we know immediately that $|\of|=192$.
   Introduce
   \eqa
   \label{def1}
    \eta=\gamma t\gamma t^2\gamma;\\
   \label{def2}
    \alpha=(t^2\gamma)^2, \beta=t\gamma t^2\gamma t;\\
   \label{def3}
    x=e_1\eta, y= e_4\eta, q=e_2\eta
   \ena
   then the structure of $\of$ can be summarized as
   \eqa
   \label{so4g1}
    x^2=y^2=q^4=e,yx=xy,qx=xq^3,qy=yq^3\\
   \label{so4g2}
    \alpha^2=\beta^2=t^3=e,
    \beta\alpha=\alpha\beta,
    t\alpha=\alpha\beta t, t\beta=\alpha t\\
   \label{so4g3}
    \alpha x=q\beta,\alpha y=q^3\beta,\alpha q=x\beta\\
   \label{so4g4}
    \beta x=q^3\alpha, \beta y=q\alpha,\beta q=y\alpha\\
   \label{so4g5}
    tx=xt^2,ty=q^3t^2,tq=yt^2
   \ena
   together with
   \eq
   \label{so4g6}
    x\alpha=\beta q^3, xt=t^2x; y\beta=\alpha q^3, yt=t^2q^3;
    qt=t^2 y
   \en
   Accordingly, each group element can be expressed as a ``normal ordering" product of $x,y\rightarrow q\rightarrow
   \alpha,\beta\rightarrow t$ and their powers from left to right.
   Throwing away all classes which belong to $\ohf$ but not to $\of$,
   there are 11 left which are $1,3,5,7,8,11,12,14,15,18,20$ in Table \ref{tab1}. The $14$th and the $20$th will part
   into two classes with equal numbers of elements each under adjoint action of
   $\of$ which are denoted as $14, 14^\pr, 20, 20^\pr$.
   Therefore, there are 13 \conv classes in $\of$.\\
   \\
   Due to the fact that $\ofd\lhd \ohfd$,
   the diagram
   \[
    \CDalign{0&\CDto&Z_2  &\CDto &\ohfd &\CDto^{\pi}&\ohf &\CDto&1 \cr
              &     &\CDeq&      &\CDup &           &\CDup&        \cr
             0&\CDto&Z_2  &\CDto &\ofd  &\CDto      &\of  &\CDto&1  }
   \]
   is commutative. We can lift generating relations
   (\ref{so4g1})(\ref{so4g2})...(\ref{so4g6})
   to
   \eqann
    x^2=y^2=q^4=-1,yx=-xy,qx=-xq^3,qy=-yq^3\\
    \alpha^2=\beta^2=t^3=-1,
    \beta\alpha=-\alpha\beta,
    t\alpha=-\alpha\beta t, t\beta=\alpha t\\
    \alpha x=q\beta,\alpha y=q^3\beta,\alpha q=x\beta\\
    \beta x=q^3\alpha, \beta y=q\alpha,\beta q=y\alpha\\
    tx=-xt^2,ty=q^3t^2,tq=yt^2\\
    x\alpha=\beta q^3, xt=-t^2x; y\beta=\alpha q^3, yt=t^2q^3;
    qt=t^2 y
   \enann
   by the definition (\ref{def1})(\ref{def2})(\ref{def3}).
   As to the \conv \clsf, $1,5,8,11,14,15,18,20,14^\pr,20^\pr$ violate relation $-g\sim g$,
   so that $\ofd$ is partitioned into 23 classes, suggesting that there are altogether 23
   \alrepv in which 13 \rep s are single-valued to $\of$.
   We summarize the \conv classes of $\ofd$ in Table \ref{tab3s}.
   \footnotesize
   \begin{table}[d]
   \caption{Conjugate Classes of $\ofd$}
   \label{tab3s}
   \begin{tabular}{|l|r|r|r|r|r|r|r|r|r|r|r|r|r|r|r|r|r|r|r|r|r|r|r|}\hline
     No.&$1$&$\bar{1}$&3&$5$&$\bar{5}$&7&$8$&$\bar{8}$&$11$&$\bar{11}$&12&$14$&$\bar{14}$&$14^\pr$&$\bar{14^\pr}$
     &$15$&$\bar{15}$&$18$&$\bar{18}$&$20$&$\bar{20}$&$20^\pr$&$\bar{20^\pr}$\\ \hline
     num.&1&1&12&1&1&48&12&12&12&12&24&6&6&6&6&32&32&32&32&24&24&24&24\\ \hline
     ord.&1&2&4&2&2&4&8&8&8&8&4&4&4&4&4&6&3&6&6&8&8&8&8\\ \hline
   \end{tabular}
   \end{table}
   \normalsize
  \section{\Repv theory of $\ofd$}\label{so4rep}
   All \alrepv of $\ofd$ can be reduced from those of $\ohfd$.
   Table \ref{tab4s} gives the characters of all \alrepv of $\ohfd$, respect to the classes of
   $\ofd$.
   \begin{table}[d]
   \caption{Character Table of $\ohfd$ (with respect to the classes of $\ofd$)}
   \label{tab4s}
   \tabcolsep 2pt
   \small{
   \begin{tabular}{|l|c|c|c|c|c|c|c|c|c|c|c|c|c|c|c|c|c|c|c|c|c|c|c|c|}\hline
    &1&$\bar{1}$&3&5&$\bar{5}$&7&8&$\bar{8}$&11&$\bar{11}$&12&14&$\bar{14}$
    &$14^\pr$&$\bar{14^\pr}$&15&$\bar{15}$&18&$\bar{18}$&20&$\bar{20}$&$20^\pr$
    &$\bar{20^\pr}$&$(\chi,\chi)$\\\hline
   $\chi_1^{(1)}$&1&1&1&1&1&1&1&1&1&1&1&1&1&1&1&1&1&1&1&1&1&1&1&{\bf 1}\\\hline
   $\chi_2^{(1)}$&1&1&1&1&1&$-1$&$-1$&$-1$&$-1$&$-1$
    &1&1&1&1&1&1&1&1&1&$-1$&$-1$&$-1$&$-1$&{\bf 1}\\\hline
   $\chi_1^{(2)}$&2&2&2&2&2&0&0&0&0&0&2&2&2&2&2&$-1$&$-1$&$-1$&$-1$&0&0&0&0&{\bf 1}\\\hline
   $\chi_1^{(3)}$&3&3&3&3&3&1&1&1&1&1&$-1$&$-1$&$-1$&$-1$&$-1$
    &0&0&0&0&$-1$&$-1$&$-1$&$-1$&{\bf 1}\\\hline
   $\chi_2^{(3)}$&3&3&3&3&3&$-1$&$-1$&$-1$&$-1$&$-1$&$-1$&$-1$&$-1$&$-1$&$-1$
    &0&0&0&0&1&1&1&1&{\bf 1}\\\hline
   $\chi_3^{(1)}$&1&1&1&1&1&$-1$&$-1$&$-1$&$-1$&$-1$
    &1&1&1&1&1&1&1&1&1&$-1$&$-1$&$-1$&$-1$&{\bf 1}\\\hline
   $\chi_4^{(1)}$&1&1&1&1&1&1&1&1&1&1&1&1&1&1&1&1&1&1&1&1&1&1&1&{\bf 1}\\\hline
   $\chi_2^{(2)}$&2&2&2&2&2&0&0&0&0&0&2&2&2&2&2&$-1$&$-1$&$-1$&$-1$&0&0&0&0&{\bf 1}\\\hline
   $\chi_3^{(3)}$&3&3&3&3&3&$-1$&$-1$&$-1$&$-1$&$-1$&$-1$&$-1$&$-1$&$-1$&$-1$
    &0&0&0&0&1&1&1&1&{\bf 1}\\\hline
   $\chi_4^{(3)}$&3&3&3&3&3&1&1&1&1&1&$-1$&$-1$&$-1$&$-1$&$-1$
    &0&0&0&0&$-1$&$-1$&$-1$&$-1$&{\bf 1}\\\hline
   $\chi_1^{(4)}$&4&4&0&$-4$&$-4$&0&2&2&$-2$&$-2$&0&0&0&0&0&1&1&$-1$&$-1$
    &0&0&0&0&{\bf 1}\\\hline
   $\chi_2^{(4)}$&4&4&0&$-4$&$-4$&0&$-2$&$-2$&2&2&0&0&0&0&0&1&1&$-1$&$-1$
    &0&0&0&0&{\bf 1}\\\hline
   $\chi_1^{(8)}$&8&8&0&$-8$&$-8$&0&0&0&0&0&0&0&0&0&0&$-1$&$-1$&1&1
    &0&0&0&0&{\bf 1}\\\hline
   $\chi_3^{(4)}$&4&4&0&$-4$&$-4$&0&$-2$&$-2$&2&2&0&0&0&0&0&1&1&$-1$&$-1$
    &0&0&0&0&{\bf 1}\\\hline
   $\chi_4^{(4)}$&4&4&0&$-4$&$-4$&0&2&2&$-2$&$-2$&0&0&0&0&0&1&1&$-1$&$-1$
    &0&0&0&0&{\bf 1}\\\hline
   $\chi_2^{(8)}$&8&8&0&$-8$&$-8$&0&0&0&0&0&0&0&0&0&0&$-1$&$-1$&1&1
    &0&0&0&0&{\bf 1}\\\hline
   $\chi_1^{(6)}$&6&6&$-2$&6&6&0&0&0&0&0&2&$-2$&$-2$&$-2$&$-2$&0&0&0&0
    &0&0&0&0&{\bf 1}\\\hline
   $\chi_2^{(6)}$&6&6&$-2$&6&6&2&-2&-2&-2&-2&-2&2&2&2&2&0&0&0&0
    &0&0&0&0&{\bf 2}\\\hline
   $\chi_3^{(6)}$&6&6&$-2$&6&6&0&0&0&0&0&2&$-2$&$-2$&$-2$&$-2$&0&0&0&0
    &0&0&0&0&{\bf 1}\\\hline
   $\chi_4^{(6)}$&6&6&$-2$&6&6&-2&2&2&2&2&-2&2&2&2&2&0&0&0&0
    &0&0&0&0&{\bf 2}\\\hline
   $\chi_1^{(\b{4})}$&4&-4&0&0&0&0&\ntsqt&\mtsqt&0&0&0&2&-2&2&-2
    &2&-2&0&0&\msqt&\nsqt&\msqt&\nsqt&{\bf 2}\\\hline
   $\chi_2^{(\b{4})}$&4&-4&0&0&0&0&\mtsqt&\ntsqt&0&0&0&2&-2&2&-2
    &2&-2&0&0&\nsqt&\msqt&\nsqt&\msqt&{\bf 2}\\\hline
   $\chi^{(\b{8})}$&8&-8&0&0&0&0&0&0&0&0&0&4&-4&4&-4
    &-2&2&0&0&0&0&0&0&{\bf 2}\\\hline
   $\chi_1^{(\b{12})}$&12&-12&0&0&0&0&\ntsqt&\mtsqt&0&0&0&-2&2&-2&2
    &0&0&0&0&\nsqt&\msqt&\nsqt&\msqt&{\bf 2}\\\hline
   $\chi_2^{(\b{12})}$&12&-12&0&0&0&0&\mtsqt&\ntsqt&0&0&0&-2&2&-2&2
    &0&0&0&0&\msqt&\nsqt&\msqt&\nsqt&{\bf 2}\\\hline
   \end{tabular}}
   \normalsize\\
   \\
   $(\chi,\chi)$ evaluates the inner product of a character with
   itself.
   \end{table}
   A brief observation gives some important information. Firstly, $1_1\cong 1_4,1_2\cong 1_3,2_1\cong 2_2,3_1\cong 3_4,
   3_2\cong 3_3, 4_1\cong 4_4,4_2\cong 4_3,8_1\cong 8_2,6_1\cong 6_3$. Secondly,
   omitting equivalence, $1_1$, $1_2$, $2_1$, $3_1$, $3_1$, $4_1$, $4_2$, $8_1$, $6_1$,
   remain irreducible within $\ofd$, while other 7 become reducible. Thirdly, as for each of
   these reducible ones, the inner product of the character with itself equals to 2, implying
   that it can be reduced to two \alrepv, thus there are 13 single-valued and 10
   spinor \rep s as we expected. Finally, it is one possible solution to Burside theorem
   that each of these seven reducible \rep s splits into 2 \alrepv
   with equal dimensions. We conjecture that it is the solution to our \repv
   theory of $\ofd$ and try to verify it below.
   \\
   \\
   Summarily speaking, there are nine single-valued \alrep s inherited from $\ohfd$
   \[
    1_1,1_2,2\equiv 2_1,3_1,3_2,4_1,4_2,8\equiv 8_1,6\equiv 6_1
   \]
   and we conjecture the splitting relations
   \eqann
    6_2, 6_4\rightarrow 3_{\alpha}, 3_{\beta}, 3_{\gamma}, 3_{\delta}\\
    \b{4}_1, \b{4}_2\rightarrow 2_{\alpha}, 2_{\beta}, 2_{\gamma},2_{\delta}\\
    \b{8}\rightarrow 4_{\alpha}, 4_{\beta}\\
    \b{12}_1, \b{12}_2\rightarrow 6_{\alpha}, 6_{\beta}, 6_{\gamma}, 6_{\delta}
   \enann
   \subsection{Hidden single-valued \rep s}
    The \repv matrices of $x$, $y$, $q$ in $6_2$ are written as
    \[
     x\mapsto
      \left(
       \begin{array}{cccccc}
        1&0&0&0&0&0\\
        0&1&0&0&0&0\\
        0&0&0&1&0&0\\
        0&0&1&0&0&0\\
        0&0&0&0&0&-1\\
        0&0&0&0&-1&0
       \end{array}
      \right),
     y\mapsto
      \left(
       \begin{array}{cccccc}
        1&0&0&0&0&0\\
        0&1&0&0&0&0\\
        0&0&0&-1&0&0\\
        0&0&-1&0&0&0\\
        0&0&0&0&0&1\\
        0&0&0&0&1&0
       \end{array}
      \right),
     q\mapsto
      \left(
       \begin{array}{cccccc}
        -1&0&0&0&0&0\\
        0&-1&0&0&0&0\\
        0&0&0&-1&0&0\\
        0&0&1&0&0&0\\
        0&0&0&0&0&1\\
        0&0&0&0&-1&0
       \end{array}
      \right)
   \]
    The textures of these matrices inspire us to such a hypotheses that in
    $3_{\alpha,\beta,\gamma,\delta}$, x,y,q take on a form like
    \[
      x,y\mapsto
       \left(
        \begin{array}{cc}
         \pm 1&\\&\pm H
        \end{array}
       \right),
      q\mapsto
       \left(
        \begin{array}{cc}
         \pm 1&\\&\pm Q
        \end{array}
       \right)
    \]
    where $H\equiv
       \left(
        \begin{array}{cc}
         0&1\\1&0
        \end{array}
       \right),
      Q\equiv
       \left(
        \begin{array}{cc}
         0&-1\\1&0
        \end{array}
       \right)$.
    After taking account of the \conv equivalence, only four possibilities survive
    from the totally 64, i.e.
    \eq
    \label{I}
     I:
       x\rightarrow
        \left(
         \begin{array}{cc}
          1&\\&H
         \end{array}
        \right),
       y\rightarrow
        \left(
         \begin{array}{cc}
          1&\\&-H
         \end{array}
        \right),
       q\rightarrow
        \left(
         \begin{array}{cc}
          -1&\\&Q
         \end{array}
        \right)
    \en
    \eq
    \label{II}
     II:
       x\rightarrow
        \left(
         \begin{array}{cc}
          -1&\\&H
         \end{array}
        \right),
       y\rightarrow
        \left(
         \begin{array}{cc}
          -1&\\&-H
         \end{array}
        \right),
       q\rightarrow
        \left(
         \begin{array}{cc}
          1&\\&Q
         \end{array}
        \right)
    \en
    \eq
    \label{III}
     III:
       x\rightarrow
        \left(
         \begin{array}{cc}
          1&\\&-H
         \end{array}
        \right),
       y\rightarrow
        \left(
         \begin{array}{cc}
          1&\\&H
         \end{array}
        \right),
       q\rightarrow
        \left(
         \begin{array}{cc}
          -1&\\&Q
         \end{array}
        \right)
    \en
    \eq
    \label{IV}
     IV:
       x\rightarrow
        \left(
         \begin{array}{cc}
          -1&\\&-H
         \end{array}
        \right),
       y\rightarrow
        \left(
         \begin{array}{cc}
          -1&\\&H
         \end{array}
        \right),
       q\rightarrow
        \left(
         \begin{array}{cc}
          1&\\&Q
         \end{array}
        \right)
    \en
    which also satisfy Eq.(\ref{so4g1}).
    Then we regard $\alpha,\beta,t$ as unknowns, (\ref{so4g2})..(\ref{so4g6}) as constraint,
    and solve these matrix equations. Modulo similarity, each of Eqs.(\ref{I})(\ref{II})
    (\ref{III})(\ref{IV}) gives two solutions,
    labeled as $I$,$I'$,$II$,$II'$,$III$,$III'$,$IV$,$IV'$;
    however, there is no difficulty to find out that $I\cong III$, $I'\cong III'$,
    $II\cong IV$, $II'\cong IV'$. Thus $I$,$I'$,$II$,$II'$ are what we need.
    \eqann
     3_{\alpha}\equiv I:
      \alpha\rightarrow diag(-1,1,-1),
      \beta\rightarrow diag(1,-1,-1),
      t\rightarrow
       \left(
        \begin{array}{ccc}
         0&0&1\\1&0&0\\0&1&0
        \end{array}
       \right)\\
     3_{\beta}\equiv I':
      \alpha\rightarrow diag(1,-1,-1),
      \beta\rightarrow diag(-1,-1,1),
      t\rightarrow
       \left(
        \begin{array}{ccc}
         0&0&1\\1&0&0\\0&1&0
        \end{array}
       \right)\\
     3_{\gamma}\equiv II:
      \alpha\rightarrow diag(-1,1,-1),
      \beta\rightarrow diag(1,-1,-1),
      t\rightarrow
       \left(
        \begin{array}{ccc}
         0&0&1\\-1&0&0\\0&-1&0
        \end{array}
       \right)\\
     3_{\delta}\equiv II':
      \alpha\rightarrow diag(1,-1,-1),
      \beta\rightarrow diag(-1,-1,1),
      t\rightarrow
       \left(
        \begin{array}{ccc}
         0&0&1\\-1&0&0\\0&-1&0
        \end{array}
       \right)
    \enann
   \subsection{Spinor \rep s}
    It is more straightforward to reduce out the spinor \rep s. We
    recall that the spinor \repv matrices of
    $\ohf$ are of the form of tensor product
    \[
     S_i(g)=S(g)\otimes s_i(g),
     \forall g\in\ohfd, i=\b{4}_1,\b{4}_2,\b{8},\b{12}_1,\b{12}_2
    \]
    where $S$ is given by the algebraic isomorphism from
    $Cl(E^4)$ to $M_2(\quaternion)$ and $s_i$ has the same texture (zero matrix element positions) of
    \irrv \repv $i$ of $\per{4}$.
    Additionally, for $g$ in $\ofd$, $S(g)$ takes on a 2-by-2
    block diagonal form
    \eq
     S(g)=\left(
       \begin{array}{cc}
        S_{up}(g)&0\\0&S_{down}(g)
       \end{array}
      \right)
    \en
    So it is just what we want
    \eqa
     S_{up}(x)= {1\over\sqt}\cdot
       \left(
        \begin{array}{cc}
         \cir{3\over 4}&\cir{-3\over 4}\\ \cir{1\over 4}&\cir{-1\over 4}
        \end{array}
       \right),
     S_{up}(y)={1\over\sqt}\cdot
       \left(
        \begin{array}{cc}
         \cir{3\over 4}&\cir{1\over 4}\\ \cir{-3\over 4}&\cir{-1\over 4}
        \end{array}
       \right)\\
      S_{up}(q)={1\over\sqt}\cdot
       \left(
        \begin{array}{cc}
         \cir{1\over 4}&\cir{3\over 4}\\ \cir{3\over 4}&\cir{1\over 4}
        \end{array}
       \right),
      S_{up}(t)={1\over\sqt}\cdot
       \left(
        \begin{array}{cc}
         1&-i\\1&i
        \end{array}
       \right)\\
      S_{up}(\alpha)=
       \left(
        \begin{array}{cc}
         0&-i\\-i&0
        \end{array}
       \right),
      S_{up}(\beta)=
       \left(
        \begin{array}{cc}
         -i&0\\0&i
        \end{array}
       \right)\\
     S_{down}(x)={1\over\sqt}\cdot
       \left(
        \begin{array}{ccc}
         \cir{3\over 4}&\cir{-3\over 4}\\ \cir{1\over 4}&\cir{-1\over 4}
        \end{array}
       \right),
     S_{down}(y)\rightarrow {1\over\sqt}\cdot
       \left(
        \begin{array}{ccc}
         \cir{-1\over 4}&\cir{-3\over 4}\\ \cir{1\over 4}&\cir{3\over 4}
        \end{array}
       \right)\\
     S_{down}(q)={1\over\sqt}\cdot
       \left(
        \begin{array}{ccc}
         \cir{1\over 4}&\cir{3\over 4}\\ \cir{3\over 4}&\cir{1\over 4}
        \end{array}
       \right),
     S_{down}(t)={1\over\sqt}\cdot
       \left(
        \begin{array}{cc}
         i&1\\-i&1
        \end{array}
       \right)\\
     S_{down}(\alpha)=
       \left(
        \begin{array}{cc}
         0&1\\-1&0
        \end{array}
       \right),
     S_{down}(\beta)=
       \left(
        \begin{array}{cc}
         0&-i\\-i&0
        \end{array}
       \right)
    \ena
    Keeping the second factor
    unchanged, each spinor \repv in $\ohfd$ splits into two spinor \rep s in $\ofd$,
    denoted as $2_\alpha$,$2_\beta$,$2_\gamma$,$2_\delta$,$4_\alpha$,$4_\beta$,
    $6_\alpha$,$6_\beta$,$6_\gamma$,$6_\delta$.\\
    \\
    In fact, $S(g)$ falls in the
    so-called ``chiral"-\repv of $Cl(E^4)$ in physical language.
    Due to $Cl(V)=Cl(V)_e\oplus Cl(V)_o$, and the choice of chiral-\rep, there are
    \[
     Cl(E^4)_e\cong
       \left(
       \begin{array}{cc}
        {\bf H}&O\\0&{\bf H}
       \end{array}
      \right),
     Cl(E^4)_o\cong
       \left(
       \begin{array}{cc}
        0&{\bf H}\\{\bf H}&0
       \end{array}
      \right)
    \]
    Notice that $\ofd<Spin(4)\subset Cl(E^4)_e$, so our
    reducing process for spinor \rep s roots in the structure of Clifford
    algebra.\\

   Conclusively, our conjecture gives all \alrep s of $\overline{SO_4}$ whose characters are summarized in Table \ref{tab5s}.\\
     \footnotesize
   \tabcolsep 3pt
   \begin{table}[d]
   \caption{Character Table of $\ofd$}
   \label{tab5s}
    \begin{tabular}{|l||c|c|c|c|c|c|c|c|c|c|c|c|c|c|c|c|c|c|c|c|c|c|c|}\hline
     &$1$&$\bar{1}$&$3$&$5$&$\bar{5}$&$7$&8&$\bar{8}$&$11$&$\bar{11}$&$12$&$14$&$\bar{14}$
     &$14^{'}$&$\bar{14^{'}}$&$15$&$\bar{15}$&$18$&$\bar{18}$&$20$&$\bar{20}$&$20^{'}$
     &$\bar{20^{'}}$\\\hline
     num.&1&1&12&1&1&48&12&12&12&12&24&6&6&6&6&32&32&32&32&24&24&24&24\\
     \hline\hline
     $\chi_1^{(1)}$&1&1&1&1&1&1&1&1&1&1&1&1&1&1&1&1&1&1&1&1&1&1&1\\\hline
     $\chi_2^{(1)}$&1&1&1&1&1&$-1$&$-1$&$-1$&$-1$&$-1$
      &1&1&1&1&1&1&1&1&1&$-1$&$-1$&$-1$&$-1$\\\hline
     $\chi^{(2)}$&2&2&2&2&2&0&0&0&0&0&2&2&2&2&2&$-1$&$-1$&$-1$&$-1$&0&0&0&0\\\hline
     $\chi_1^{(3)}$&3&3&3&3&3&1&1&1&1&1&$-1$&$-1$&$-1$&$-1$&$-1$
      &0&0&0&0&$-1$&$-1$&$-1$&$-1$\\\hline
     $\chi_2^{(3)}$&3&3&3&3&3&$-1$&$-1$&$-1$&$-1$&$-1$&$-1$&$-1$&$-1$&$-1$&$-1$
      &0&0&0&0&1&1&1&1\\\hline
     $\chi_1^{(4)}$&4&4&0&$-4$&$-4$&0&2&2&$-2$&$-2$&0&0&0&0&0&1&1&$-1$&$-1$
      &0&0&0&0\\\hline
     $\chi_2^{(4)}$&4&4&0&$-4$&$-4$&0&$-2$&$-2$&2&2&0&0&0&0&0&1&1&$-1$&$-1$
      &0&0&0&0\\\hline
     $\chi^{(8)}$&8&8&0&$-8$&$-8$&0&0&0&0&0&0&0&0&0&0&$-1$&$-1$&1&1
      &0&0&0&0\\\hline
     $\chi^{(6)}$&6&6&$-2$&6&6&0&0&0&0&0&2&$-2$&$-2$&$-2$&$-2$&0&0&0&0
      &0&0&0&0\\\hline
     $\chi_{\alpha}^{(3)}$&3&3&-1&3&3&1&-1&-1&-1&-1&-1&-1&-1&3&3
      &0&0&0&0&-1&-1&1&1\\\hline
     $\chi_{\beta}^{(3)}$&3&3&-1&3&3&1&-1&-1&-1&-1&-1&3&3&-1&-1
      &0&0&0&0&1&1&-1&-1\\\hline
     $\chi_{\gamma}^{(3)}$&3&3&-1&3&3&-1&1&1&1&1&-1&-1&-1&3&3
      &0&0&0&0&1&1&-1&-1\\\hline
     $\chi_{\delta}^{(3)}$&3&3&-1&3&3&-1&1&1&1&1&-1&3&3&-1&-1
      &0&0&0&0&-1&-1&1&1\\\hline \hline
     $\chi_{\alpha}^{(2)}$&2&-2&0&2&-2&0&\nsqt&\msqt&\msqt&\nsqt&0&2&-2&0&0
      &1&-1&1&-1&0&0&\msqt&\nsqt\\\hline
     $\chi_{\beta}^{(2)}$&2&-2&0&2&-2&0&\msqt&\nsqt&\nsqt&\msqt&0&2&-2&0&0
      &1&-1&1&-1&0&0&\nsqt&\msqt\\\hline
     $\chi_{\gamma}^{(2)}$&2&-2&0&-2&2&0&\nsqt&\msqt&\nsqt&\msqt&0&0&0&-2&2
      &1&-1&-1&1&\msqt&\nsqt&0&0\\\hline
     $\chi_{\delta}^{(2)}$&2&-2&0&-2&2&0&\msqt&\nsqt&\msqt&\nsqt&0&0&0&-2&2
      &1&-1&-1&1&\nsqt&\msqt&0&0\\\hline
     $\chi_{\alpha}^{(4)}$&4&-4&0&4&-4&0&0&0&0&0&0&4&-4&0&0
      &-1&1&-1&1&0&0&0&0\\\hline
     $\chi_{\beta}^{(4)}$&4&-4&0&-4&4&0&0&0&0&0&0&0&0&-4&4
      &-1&1&1&-1&0&0&0&0\\\hline
     $\chi_{\alpha}^{(6)}$&6&-6&0&6&-6&0&\nsqt&\msqt&\msqt&\nsqt&0&-2&2&0&0
      &0&0&0&0&0&0&\nsqt&\msqt\\\hline
     $\chi_{\beta}^{(6)}$&6&-6&0&6&-6&0&\msqt&\nsqt&\nsqt&\msqt&0&-2&2&0&0
      &0&0&0&0&0&0&\msqt&\nsqt\\\hline
     $\chi_{\gamma}^{(6)}$&6&-6&0&-6&6&0&\nsqt&\msqt&\nsqt&\msqt&0&0&0&2&-2
      &0&0&0&0&\nsqt&\msqt&0&0\\\hline
     $\chi_{\delta}^{(6)}$&6&-6&0&-6&6&0&\msqt&\nsqt&\msqt&\nsqt&0&0&0&2&-2
      &0&0&0&0&\msqt&\nsqt&0&0\\\hline
    \end{tabular}
    \normalsize
   \end{table}


   {\bf Acknowledgements}\\
    This work was supported by Climb-Up (Pan Deng) Project of
    Department of Science and Technology in China, Chinese
    National Science Foundation and Doctoral Programme Foundation
    of Institution of Higher Education in China.
    One of the authors J.D. is grateful to Dr. L-G. Jin for his advice on this paper.
  \appendix
  \setcounter{equation}{0}
  \renewcommand{\theequation}{A-\arabic{equation}}
  \setcounter{definition}{0}
  \renewcommand{\thedefinition}{A-\arabic{definition}}
  \setcounter{lemma}{0}
  \renewcommand{\thelemma}{A-\arabic{lemma}}
  \setcounter{proposition}{0}
  \renewcommand{\theproposition}{A-\arabic{proposition}}
  \setcounter{corollary}{0}
  \renewcommand{\thecorollary}{A-\arabic{corollary}}
  \section{Fundamental lemma of n-dimensional Euclidean geometry}
  \label{app2}
   \begin{lemma}
   \label{fundament}
   (Weak form)
    Let $p_i, i=0,1,2,...,n$ be $n+1$ points in
    n-dimensional Euclidean space $\eun$ which are non-collinear
    and give $n+1$ non-negative real numbers $d_i, i=0,1,2,...,n$, then there exists
    at most one point $p\in \eun$ s.t. $d(p,p_i)=d_i$.
   \end{lemma}
   \prf
    Without losing generality, set $p_0=(0,0,...,0)$ and
    understand $p, p_i, i=1,2,...,n$ as vectors in $\eun$.
    Consider equation set
    \eqa
    \label{disf}
     (p-p_i, p-p_i)&=&d_i^2, i=1,2,...,n\\
    \label{disf1}
     (p,p)&=&d_0^2
    \ena
    where $(,)$ is standard inner product in $E^n$.
    Substitute (\ref{disf1}) into (\ref{disf})
    \eq
    \label{disf2}
     (p_i,p)=\hf(d_0^2-d_i^2+(p_i,p_i)), i=1,2,...,n
    \en
    The non-collinearity implies that (\ref{disf2}) has a solution
    $p$. The weak form of {\it fundamental lemma of Euclidean
    geometry} follows. \\
   \endprf
  
  \newpage
  \Large
  \begin{center}
    Table List\\
    \ref{tab1}. Conjugate Classes of $\ztn{4}\smdp\per{4}$\\
    \ref{tab2}. Conjugate Classes of $\ohd$\\
    \ref{tab3}. Character Table of $\ztf$\\
    \ref{tabch}. Character Table of $\ohd$\\
    \ref{tab3s}. Conjugate Classes of $\ofd$\\
    \ref{tab4s}. Character Table of $\ohfd$ (with respect to the classes of
    $\ofd$)\\
    \ref{tab5s}. Character Table of $\ofd$
   \end{center}
 \end{document}